\documentclass[sigconf]{acmart}

\copyrightyear{2020} 
\acmYear{2020} 
\setcopyright{acmcopyright}\acmConference[SEAMS '20]{IEEE/ACM 15th International Symposium on Software Engineering for Adaptive and Self-Managing Systems}{October 7--8, 2020}{Seoul, Republic of Korea}
\acmBooktitle{IEEE/ACM 15th International Symposium on Software Engineering for Adaptive and Self-Managing Systems (SEAMS '20), October 7--8, 2020, Seoul, Republic of Korea}
\acmPrice{15.00}
\acmDOI{10.1145/3387939.3391603}
\acmISBN{978-1-4503-7962-5/20/05}

\usepackage{amsmath,amssymb,amsfonts}
\usepackage{esvect}
\usepackage{graphicx}
\usepackage{subfigure}
\usepackage{booktabs}
\usepackage{tabularx}
\usepackage{multirow}
\usepackage{balance}
\usepackage{xcolor}
\usepackage{enumitem}
\usepackage{mathtools}
\usepackage{wrapfig}
\usepackage{ragged2e}
\usepackage[all]{nowidow}
\usepackage{stfloats}
\usepackage{anyfontsize}

\newcolumntype{Y}{>{\centering\arraybackslash}X}

\usepackage{colortbl}
\usepackage{listings}

\definecolor{javared}{rgb}{0.6,0,0} % for strings
\definecolor{javagreen}{rgb}{0.25,0.5,0.35} % comments
\definecolor{javapurple}{rgb}{0.5,0,0.35} % keywords
\definecolor{javadocblue}{rgb}{0.25,0.35,0.75} % javadoc
\definecolor{javagrey}{rgb}{0.46,0.45,0.48} % annotations

\lstdefinestyle{Alg}{
  basicstyle=\ttfamily\footnotesize,
  breaklines=true,
  tabsize=2,
  mathescape,
  numbers=left,
  xleftmargin=2.5em,
  xrightmargin=0.5em,
  frame=tb,
  framexleftmargin=2em,
  emph={Algorithm,Input,Output,for,each,do,if,else,Function,while,let,be,repeat,until,return,times,and,or,break,in,then,},
  emphstyle={\textbf},
  escapechar=?,
  morecomment=[l][\color{javagreen}]{//},
  columns=flexible,
}
\usepackage{framed}
\usepackage{tikz}
\definecolor{light-gray}{gray}{0.9}
\usepackage[framemethod=TikZ]{mdframed}
\mdfsetup{skipabove=5pt,skipbelow=3pt}
\usepackage{lipsum}
\mdfdefinestyle{RQFrame}{%
	linecolor=black,
	outerlinewidth=0.15pt,
	roundcorner=3pt,
	innertopmargin=2pt,
	innerbottommargin=2pt,
	innerrightmargin=4pt,
	innerleftmargin=4pt,
	backgroundcolor=light-gray}
\begin{document}

\title[Dynamic Adaptation of Software-defined Networks for IoT Systems: A Search-based Approach]{Dynamic Adaptation of Software-defined Networks for\texorpdfstring{\\}{ }IoT Systems: A Search-based Approach}
% Search-based Self-adaptive Network Configuration for IoT Systems
% Search-based Self-adaptive Quality Assurance for IoT Systems
% Using Search-based Self-adaptation for Maintaining Quality of IoT Services
% Using Search-based Dynamic Adaptation for Configuring an Underlying Network of IoT Systems

\author{Seung Yeob Shin}
\affiliation{%
  \institution{University of Luxembourg, Luxembourg}
}
\email{seungyeob.shin@uni.lu}

\author{Shiva Nejati}
\affiliation{%
  \institution{University of Ottawa, Canada}
}
\affiliation{%
  \institution{University of Luxembourg, Luxembourg}
}
\email{snejati@uottawa.ca}

\author{Mehrdad Sabetzadeh}
\affiliation{%
  \institution{University of Ottawa, Canada}
}
\affiliation{%
  \institution{University of Luxembourg, Luxembourg}
}
\email{msabetza@uottawa.ca}

\author{Lionel C. Briand}
\affiliation{%
  \institution{University of Ottawa, Canada}
}
\affiliation{%
  \institution{University of Luxembourg, Luxembourg}
}
\email{lbriand@uottawa.ca}

\author{Chetan Arora}
\affiliation{%
  \institution{SES Networks, Luxembourg}
}
\affiliation{%
  \institution{University of Luxembourg, Luxembourg}
}
\affiliation{%
  \institution{Deakin University, Australia}
}
%\email{chetan.arora@ses.com}
\email{chetan.arora@deakin.edu.au}

\author{Frank Zimmer}
\affiliation{%
  \institution{SES Networks, Luxembourg}
}
\email{frank.zimmer@ses.com}

%\renewcommand{\shortauthors}{Shin, Nejati, Sabetzadeh, Briand, Arora, and Zimmer}

% !TEX root =  paper.tex

\begin{abstract}

The concept of Internet of Things (IoT) has led to the development of many complex and critical systems such as smart emergency management systems. IoT-enabled applications typically depend on a communication network for transmitting large volumes of data in unpredictable and changing environments. These networks are prone to congestion when there is a burst in demand, e.g., as an emergency situation is unfolding, and therefore rely on configurable software-defined networks (SDN). In this paper, we propose a dynamic adaptive SDN configuration approach for IoT systems. The approach enables resolving congestion in real time while minimizing network utilization, data transmission delays and adaptation costs. Our approach builds on existing work in dynamic adaptive search-based software engineering (SBSE) to reconfigure an SDN while simultaneously ensuring multiple quality of service criteria. We evaluate our approach on an industrial national emergency management system, which is aimed at detecting disasters and  emergencies, and facilitating recovery and rescue operations by  providing first responders with a reliable communication infrastructure. Our results indicate that (1)~our approach is able to efficiently and effectively adapt an SDN to dynamically resolve congestion, and (2)~compared to two baseline data forwarding algorithms that are static and non-adaptive, our approach increases data transmission rate by a factor of at least 3 and decreases data loss \hbox{by at least 70\%.}
\end{abstract}

%\begin{CCSXML}
%<ccs2012>
%  <concept_id>10010520.10010553.10010562</concept_id>
%  <concept_desc>Computer systems organization~Embedded systems</concept_desc>
%  <concept_significance>500</concept_significance>
%  <concept_id>10010520.10010575.10010755</concept_id>
%  <concept_desc>Computer systems organization~Redundancy</concept_desc>
%  <concept_significance>300</concept_significance>
%  <concept_id>10010520.10010553.10010554</concept_id>
%  <concept_desc>Computer systems organization~Robotics</concept_desc>
%  <concept_significance>100</concept_significance>
%  <concept_id>10003033.10003083.10003095</concept_id>
%  <concept_desc>Networks~Network reliability</concept_desc>
%  <concept_significance>100</concept_significance>
%</ccs2012>
%\end{CCSXML}
%
%\ccsdesc[500]{Computer systems organization~Embedded systems}


\maketitle

% !TEX root =  paper.tex

\section{Introduction}
\label{sec:intro}

\begin{sloppypar}
The recent proliferation of sensors, actuators and inexpensive network-enabled devices in homes, workplaces, public spaces and nature  provides several opportunities to build intelligent systems that can improve our lives in many different ways.  These devices, when used in combination with wired and wireless connectivity, have created a surge of interest in the concept of Internet of Things (IoT). Systems enabled by IoT perform a task by connecting sensors and actuators and many previously unconnected things through the Internet~\cite{Al-Fuqaha:15,Atzori:10}. A notable example of an IoT-enabled system is an emergency management system that monitors a large geographical area through a network of sensors to detect potential disasters (e.g., fire, floods, hurricanes, earthquakes) as early as possible and to provide a communication platform between the responsible organizations and people to enable quick action and minimize loss of life and damages.
\end{sloppypar}

Successful IoT systems necessarily depend on an underlying communication system that can transmit large volumes of data in an efficient, effective and flexible way. Such a communication system should, in particular, be able to adapt to changes in the environment and maintain a reasonable quality of service when, for example, the traffic for a particular network route increases dramatically due to a massive  demand from system users. Recently, software-defined networks (SDN)~\cite{SDN:15} have started to enable such flexible and effective communication systems. The idea behind SDN is to transfer the control of networks from localized fixed-behavior controllers distributed over a set of switches to a centralized and programmable software controller that can react to environment changes in a timely fashion by efficiently reconfiguring the entire network. With software being an integral part of SDN, developing network controllers needs interdisciplinary considerations which include not only network engineering, \hbox{but also \emph{software engineering}~\cite{Lopes:16}.}

For an IoT system that builds on SDN, the controller is responsible for ensuring that the network is configured in such a way as to maintain the quality of service at a desired level. In this paper, we focus on developing effective reconfiguration techniques for SDN to improve the quality of service in IoT systems. Such techniques should be able to continuously monitor environment changes and dynamically reconfigure the system accordingly in order to optimize multiple quality of service criteria such as minimizing data loss, communication delays and reconfiguration costs. There are a number of existing research threads on ensuring the quality of service for traditional networks~\cite{Mathis:96,Alizadeh:10,Ferlin:16,He:16,Betzler:16}. Some more recent approaches study dynamic reconfiguration of SDN to maximize quality of service~\cite{Chiang:18,Gay:17,Huang:16}. None of these lines of work, however, consider or optimize the configuration of an SDN for \emph{multiple} quality of service criteria simultaneously. The problem of configuration for the purpose of optimizing multiple criteria has been studied in prior research threads for design-time software development~\cite{Zoghi:16,Andrade:13,Ramirez:09}. These studies, however, are geared toward offline optimization of system design or architecture, and cannot address the challenge of online and dynamic SDN reconfiguration.

\begin{figure}[t]
	\centerline{\includegraphics[width=0.75\columnwidth]{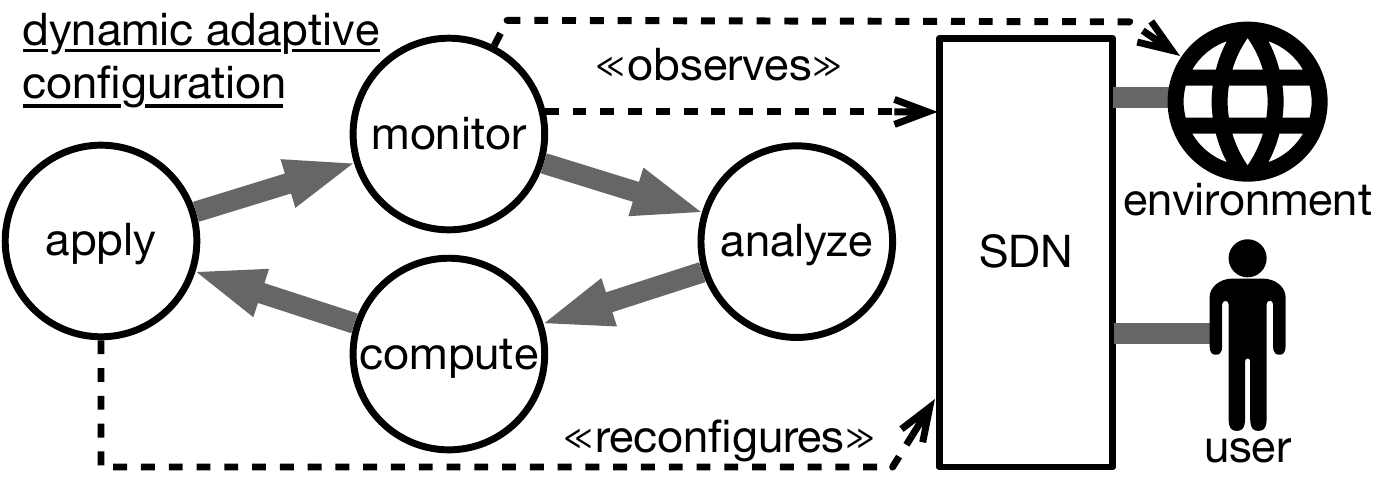}}
	\vspace{-1.0em}
	\caption{An overview of our \emph{Dynamic adaptIve CongEstion control algorithm for SDN (DICES)}.}
	\Description{An overview of our \emph{Dynamic adaptIve CongEstion control algorithm for SDN (DICES)}.}
	\label{fig:overview}
	\vspace{-1.5em}
\end{figure}

In this paper, we propose a dynamic adaptive configuration technique to resolve congestion in SDN in an online manner, while  minimizing data transmission delays and reconfiguration costs. We refer to our approach as \emph{Dynamic adaptIve CongEstion control algorithm for SDN (DICES)}. Inspired by feedback-loop control systems~\cite{Kephart:03}, DICES realizes the control loop shown in Fig.~\ref{fig:overview} and consisting of the following steps: (1)~\emph{monitor} the SDN to collect network information,  (2)~\emph{analyze} the network to determine whether it is congested, (3)~\emph{compute} a reconfiguration if congestion is detected, and (4)~\emph{apply} the new configuration to the actual SDN. The control loop is executed periodically and may reconfigure the SDN at each period if congestion is detected. The ``compute'' step of DICES uses a tailored  multi-objective search algorithm to optimize multiple quality of service criteria simultaneously. Specifically, the algorithm minimizes the following three objectives: \emph{network-link utilization}, \emph{transmission delay}  and \emph{reconfiguration cost}. In order to be executed in a real-time manner, DICES has to be efficient. Hence, instead of searching for the very best reconfiguration option, the approach aims to find good-enough solutions sufficiently quickly. Consistent with this  goal, we build on the research field of dynamic adaptive search-based software engineering (SBSE)~\cite{Harman:12} to enable the computation \hbox{component in charge of the reconfiguration of an SDN.}

DICES has to be integrated and executed together with an actual SDN control platform. We develop DICES as an application within ONOS~\cite{Berde:14} -- a widely used open-source SDN control platform. To evaluate DICES, we rely on an open-source network emulator, Mininet~\cite{Lantz:10}. Mininet enables us to create and emulate realistic virtual networks with different topologies and characteristics. Alongside Mininet, we employ an open-source network traffic flow generator, D-ITG~\cite{Botta:12}, to generate IoT traffic scenarios by combining sensor, video, audio and data streams. Our implementation of DICES can be integrated in a straightforward way into an actual network system. Nevertheless, we elect to evaluate our approach based on emulation through Mininet. This choice, which follows standard engineering practice, is motivated by two main factors: First, the large-scale and systematic experiments that we perform would be prohibitively expensive to set up using real hardware. Second, we need to evaluate our approach on varying networks with different sizes and properties. With real hardware, we would not have the required flexibility.

We assess the performance of DICES on ten synthetic and one industrial SDN. The industrial SDN is a national emergency management system in Luxembourg. The information about the topology and the IoT traffic scenarios for this SDN is provided by SES, a leading satellite operator, which is in charge of assessing the infrastructure for the national emergency management system. Our results show that: (1)~DICES efficiently and effectively adapts an SDN to resolve congestion, (2)~the execution time of DICES scales linearly with the network size and the number of traffic flows, and (3)~compared to two baseline solutions commonly used in practice~\cite{Poularakis:19,Amin:18,Bianco:17,Rego:17,Caria:15,Bianco:15,Agarwal:13}, DICES leads to data transmissions that are at least 3 times faster while reducing data loss by at least 70\%. Our case study data is available online~\cite{Artifacts}. 
%To our knowledge, DICES is the first SDN application available online which aims to address congestion while minimizing transition delays and reconfiguration costs.

\emph{Organization.} The rest of this paper is organized as follows. Section~\ref{sec:motivation} motivates the paper. Section~\ref{sec:approach} describes DICES. Section~\ref{sec:exp} evaluates DICES. Section~\ref{sec:related} compares with related work. Section~\ref{sec:conclusions} concludes this paper.
% !TEX root =  paper.tex

\section{Motivating Case Study}
\label{sec:motivation}

We motivate our work with an IoT-enabled national emergency management system, currently under study by SES, for public protection and disaster relief. We refer to this system as \emph{EMS} in the rest of the paper. EMS is responsible for generating early warnings about potential disasters,  detecting  natural or man-made emergencies, and  facilitating response/recovery operations by providing emergency workers or governmental bodies with a reliable and efficient communication and data transfer infrastructure. 

\begin{figure}[t]
	\centerline{\includegraphics[width=0.85\columnwidth]{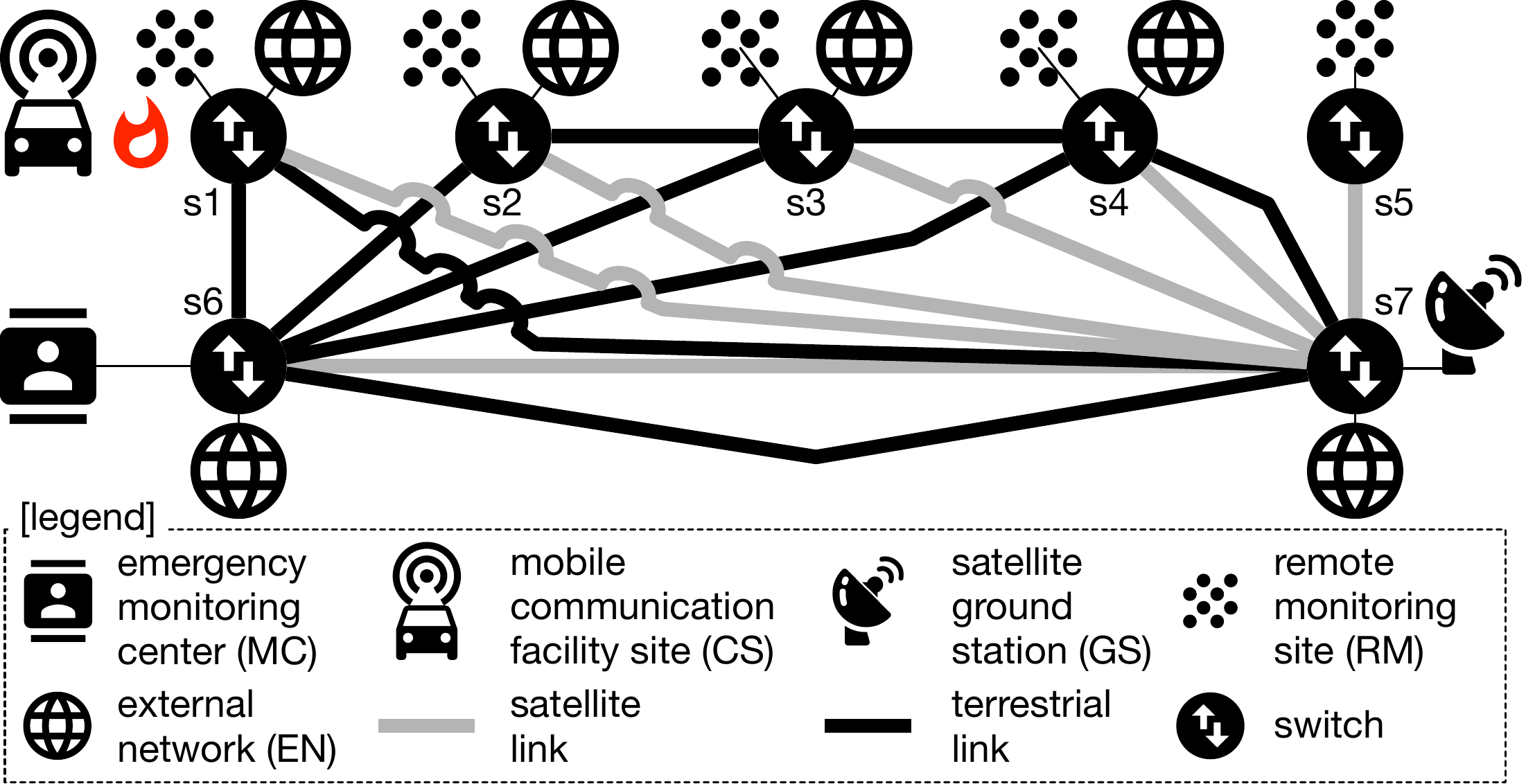}}
	\vspace{-0.7em}
	\caption{A conceptual view of an emergency management system (EMS).}
	\Description{A conceptual view of an emergency management system (EMS).}
	\label{fig:ems topo}
	\vspace{-1.5em}
\end{figure}

Fig.~\ref{fig:ems topo} shows a conceptual view of EMS for an example topology suggested by SES. EMS employs SDN to interconnect four types of sites, namely \emph{remote monitoring site}, \emph{emergency monitoring center}, \emph{satellite ground station}, and \emph{mobile communication facility site}. The interconnections are realized using SDN switches (s1--s7), terrestrial links (e.g., optical fiber links) and satellite links. The key characteristics of the four EMS sites are described below.
\\ \hspace*{.1em} $\bullet$ \emph{Remote monitoring sites (RM)} continuously monitor and gather environment data using  sensor networks. In Fig.~\ref{fig:ems topo}, switches s1--s5  are connected to remote monitoring sites. % such as the meteorological data networks
\\ \hspace*{.1em} $\bullet$ \emph{Emergency monitoring centers (MC)} control and monitor the entire EMS by aggregating data from the remote sites. They further facilitate decision making for emergency handling by controlling the entire network and by processing the aggregated data. EMS has one emergency monitoring center attached to s6, \hbox{as depicted in Fig.~\ref{fig:ems topo}.}
\\ \hspace*{.1em} $\bullet$ \emph{Satellite ground stations (GS)} are responsible for routing data streams transmitted by satellites. All satellite connections need to pass through a satellite ground station. EMS has one satellite ground station attached to s7, as shown in Fig.~\ref{fig:ems topo}.
%The data transfer via satellites from the remote monitoring sites (s1--s5) to the emergency monitoring site (s6) has to go through the satellite ground station (s7). \textcolor{red}{LB: why since there are terrestrial links?}
%Satellite connectivity adds resilience to the EMS by providing a backup to the terrestrial links which might be damaged during disasters.
\\ \hspace*{.1em} $\bullet$ \emph{Mobile communication facility sites (CS)} are used by emergency workers and first responders for communication during an actual emergency. Unlike the other EMS sites that are operational at all times, the mobile communication facility comes into play only during or after an emergency. The mobile facility site is primarily used as a communication hotspot for audio and video transmission between a remote monitoring site and the emergency monitoring center. In our case study, we assume that an emergency situation, e.g., a natural disaster, occurs in the area close to s1. Hence, in  Fig.~\ref{fig:ems topo}, the mobile communication facility site is located at s1. 

Finally, as shown in  Fig.~\ref{fig:ems topo}, the EMS network can further be connected to \emph{external (legacy) networks (EN)} to allow access to remote monitoring sites.

During an emergency, the EMS data traffic volume increases by many folds. The remote monitoring sites transmit monitored data streams to the emergency monitoring center. The mobile communication facility site and the emergency monitoring center exchange high-bandwidth demanding streams such as high definition video and audio for real-time updates. The emergency monitoring center sends earth-observation images (i.e., maps) to the mobile communication facility site in order to help plan an appropriate \hbox{recovery strategy.}

EMS is highly prone to congestion during emergencies due to the increased volume of demand. Such congestion leads to increased latency, information loss and inability to communicate with one or more sites. While such congested networks are common during emergencies, critical systems such as EMS are expected to be resilient and find ways to avoid or mitigate congestion. Failing to do so can have dire consequences. EMS is thus subject to strict quality of service requirements so that it will operate through network issues without intolerable delays or information loss. To this end, SES is interested in DICES as a way to ensure that EMS can sustain emergency situations and satisfy its quality of \hbox{service requirements.}
%and continuous changes in data transfer requests
% and interruptions 
%, on top of the facilities provided by SDN,

% !TEX root =  paper.tex

\section{Approach}
\label{sec:approach}

%\subsection{Background on SDN}
%\label{subsec:background}

%\begin{figure}[t]
%  \centerline{\includegraphics[width=0.9\columnwidth]{figs/SDN}}
%  \caption{A conceptual architecture of SDN focusing on the relevant components to our study.}
%  \label{fig:SDN}
%\end{figure}

The separation between software-defined data control and the physical aspects of network systems is a key feature of SDN~\cite{SDN:15}. The SDN architecture is composed of three layers: \emph{infrastructure}, \emph{control}, and \emph{application}. The infrastructure layer is comprised of physical entities such as links and switches that enable data flows based on forwarding rules instructed by the control layer. The control layer hosts one or multiple SDN controllers distributed across the network.  This layer is responsible for managing infrastructure entities, e.g., switches and links, based on algorithms provided by the application layer. In Section~\ref{subsec:problem}, we provide an abstract formalization of SDN concepts and use them to define the problem of \hbox{network congestion.}

%Further, when  multiple SDN controllers  are present, the control layer is responsible for synchronizing them.

%This is demonstrated via the conceptual architecture of SDN shown in Fig.~\ref{fig:SDN}. 

%SDN are able to dynamically control the network components for various complex applications, (e.g., data centers, cloud computing, emergency management systems). In fact, the control layer acts as a bridge between network applications and infrastructure entities, and it further provides a unified and global view of the entire network which may include multiple distributed controllers. Therefore, application layer developers are shielded from physical layer details and do not have to deal with complex synchronization of distributed networks. The control layer then converts the shortest paths computed by the application layer into specific  data-forwarding rules applied to the entities at the infrastructure layer.

%In this paper,  we develop an SDN application that re-routes data flows dynamically to eliminate congestions while minimizing the cost of re-routing and overall network delays.

The behavior of the control layer can be modified and extended by the application layer. Users can develop their own applications to apply domain-specific data forwarding, security or failure management algorithms. Specifically, the SDN application layer includes a data-forwarding algorithm that directs data flows between any pair of switches through  the weighted shortest path between the switches. This default data-forwarding algorithm is described in Section~\ref{subsec:sdn data forwarding}.  Since SDN controller behavior is  programmable through applications, we can enhance the data-forwarding function of SDN using DICES  as described in Section~\ref{subsec:dices}.  

\subsection{Problem Description}
\label{subsec:problem}

%For the consistent use of terms, however, this paper describes the problem by using the terminology introduced in the prior sections such as networks, switches, links, bandwidths, delays rather than graphs, vertices, edges, capacities, and lengths, respectively. 

\begin{figure}[t]
\centering
\subfigure[An example network]{%
	\includegraphics[width=0.45\columnwidth]{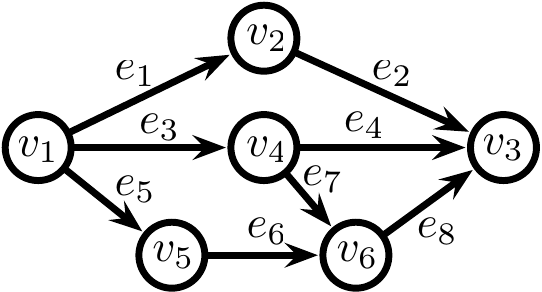}
	\label{fig:eg network}
}
\subfigure[$c(e)$ and $l(e)$ values]{
%\hspace{-8pt}
\begin{minipage}[t]{1\columnwidth}
	\centering
	\fontsize{8}{8}\selectfont
  \setlength{\tabcolsep}{0.5em} % for the horizontal padding
  \def\arraystretch{0.8}%  1 is the default, change whatever you need
  \begin{tabularx}{1.0\columnwidth}{lcccccccc}
    \toprule
    bandwidth & $c(e_1)$ & $c(e_2)$ & $c(e_3)$ & $c(e_4)$ & $c(e_5)$ & $c(e_6)$ & $c(e_7)$ & $c(e_8)$\\
    Mbps & 20 & 20 & 10 & 10 & 20 & 20 & 10 & 20\\    
    \midrule
    delay & $l(e_1)$ & $l(e_2)$ & $l(e_3)$ & $l(e_4)$ & $l(e_5)$ & $l(e_6)$ & $l(e_7)$ & $l(e_8)$\\
    ms & 250 & 250 & 250 & 250 & 25 & 25 & 25 & 25\\
    \bottomrule
  \end{tabularx}
  \vspace{0.5em}
\end{minipage}
\label{fig:eg bandwidths}
}
\vspace{-1.0em}
\caption{An example network: (a)~network topology and (b)~$c(e)$ bandwidth and $l(e)$ delay values for each link in the network topology.}
\Description{An example network: (a)~network topology and (b)~$c(e)$ bandwidth and $l(e)$ delay values for each link in the network topology.}
\vspace{-1.5em}
\end{figure}

In this section, we describe SDN topologies using directed graphs and formalize SDN traffic concepts. We then define the problem of network congestion.  We define an SDN  network as a tuple $G=(V,E,c,l)$, where $V$ is a set of switches, $E \subseteq V \times V $ is a set of directed links between switches, $c$ is a bandwidth function $c: E \rightarrow \mathbb{N}$ assigning a positive integer value $c(e)$ to every link $e \in E$, and $l$ is a delay function $l: E \rightarrow \mathbb{N}$ assigning a positive integer value $l(e)$ to every link $e \in E$. For example, Fig.~\ref{fig:eg network} presents an example SDN topology with six switches, $v_1,v_2,...,v_6$, and eight directed links $e_1,e_2,...,e_8$; and, Fig.~\ref{fig:eg bandwidths} shows the bandwidth and delay values of each link in Fig.~\ref{fig:eg network}. The network of EMS in Fig~\ref{fig:ems topo} could be represented using a graph similar to that in Fig.~\ref{fig:eg network}  where every node in Fig.~\ref{fig:eg network} corresponds to a switch in EMS and every link in  Fig.~\ref{fig:eg network} corresponds to a terrestrial or satellite link in EMS. Note that EMS terrestrial and satellite links are bidirectional and thus have to be represented as two directed graph links.

A network request $q$ specifies a data stream that should be sent by a source switch $s$  to a terminal switch $t$. Each network request $q$ has a source switch $q.s$, a terminal switch $q.t$ and a data stream of size (or bandwidth) $q.d$.  Note that $q.d$ may vary over time, but, for notational simplicity, we capture $q.d$ as a constant. We produce a different request if $q.d$ changes and remove the old one. To process each request $q$, a flow $f$ is created.  A flow describes a directed path, i.e., a sequence of links, in $G$ that is used to transmit the data stream of $q$. We denote by $f.q$ the request $q$ related to a flow $f$,  and by $f.p$ the directed path that is used to carry the data of $q$ from $q.s$ to $q.t$. Let $F$ be a set of flows. We denote by $\mathit{links}(f)$ the set of links on the directed path $f.p$ and by  $\mathit{links}(F) = \cup_{f \in F} \mathit{links}(f)$ the set of all the links of the flows in $F$. Finally, we denote the subset of flows in $F$ going through link $e$ by $\mathit{flows}(e,F)$.

%We define a request $q$ as $(\mathit{id},s,t,d)$, where $\mathit{id}$ is a unique identifier, $s \in V$ is a source switch, $t \in V$ is a terminal switch, and $d \in \mathbb{N}$ is the required flow size to carry the data stream of $q$. We write $q.s$, $q.t$, and $q.d$ to refer to  the source $s$, the terminal $t$, and the bandwidth $d$ of a request $q$, respectively. 
 %Let $Q$ be a set of requests. \textcolor{red}{A question that may arise here is whether the requests are time-stamped or not (i.e., queue vs. set).}A flow $f$ is created to process a request $q$. It describes a directed path, i.e., a sequence of links, in $G$ that is used to transmit the data stream of $q$. We define a flow $f$ by a tuple $(q, p)$, where $q$ is a request being processed by the flow, and $p$ is a directed path from $q.s$ to $q.t$. We write  $f.p$ and $f.q$ to refer to the path $p$ and the request $q$ of  flow $f$, respectively. 
%denote the request $q$ being processed and the directed path $p$ of a flow $f$ by $f.q$ and $f.p$, respectively. % Let $F$ be a set of flows.

%Given the above definitions regarding networks, requests, and flows, we define some additional notation -- $\mathit{links}(f)$, $\mathit{links}(F)$, and $\mathit{flows}(e,F)$ -- to describe the problem of interest. 

%and define it as  $\mathit{flows}(e,F) = \{f~|~f \in F \wedge e \in \mathit{links}(f)\}$.

\begin{sloppypar}
The bandwidth $c(e)$ of a network link $e$ is a (limited) resource shared by different flows. A flow $f$ going through a link $e$ consumes the link's bandwidth $c(e)$ by the flow size $f.q.d$. Hence, the total size of flows going though  $e$, i.e., the throughput of $e$, should be less than or equal to the bandwidth $c(e)$. Given a set $F$ of flows, we define the throughput of $e$ for $F$ as follows: \hbox{$\mathit{throughput}(e,F) = \underset{f \in \mathit{flows}(e,F)}{\sum}f.q.d$.}
\end{sloppypar}

%\begin{small}
%\begin{equation}
%\mathit{throughput}(e,F) = \underset{f \in \mathit{flows}(e,F)}{\sum}f.q.d
%\label{eq:throughput}
%\end{equation}
%\end{small}

We say a network $G$ is congested by a given set $F$ of flows if there is some link $e$ such that $\mathit{throughput}(e,F) > c(e)$. Given a network $G$ congested  by the set $F$ of flows, we address the problem of network congestion by finding a new set  $F^a = \{f_1^a, f_2^a, ..., f_n^a\}$ of flows where (1)~each $f_j^a$ processes the same request as that  of the flow $f_j \in F$, i.e., $f_j^a.q = f_j.q$, and thus $F^a$ and $F$ have the same cardinality, i.e., $|F^a| = |F|$, and (2) $G$ is not congested by $F^a$, i.e.,  $\mathit{throughput}(e,F^a) \le c(e)$ for all $e \in \mathit{links}(F^a)$. The problem of resolving network congestion  is NP-hard~\cite{Amiri:18,Brandt:16}. Note that $F^a$ may not exist when, for example, all the links in $G$ are overutilized by network requests. In this case, we aim to compute  $F^a$ such that the maximum link throughput is minimized even if it is still congested (see Section~\ref{subsec:dices}).

%Note that the problem of interest accounts for the time-varying characteristics of a network such as time-varying arrivals of requests. Without loss of generality, we defined the problem without considering time. Given a set of sampled time $\mathbb{T} = \{0, \varDelta, ..., k{\cdot}\varDelta\}$, where $k+1$ is a number of sampled time and $\varDelta$ is a sampling time interval, however, we can denote a set of requests being processed by flows at $t \in \mathbb{T}$ by $Q_t$ and a set of flows corresponding to $Q_t$ by $F_t$. Thus, the problem definition can describe the congestion problem at time $t$ based on $Q_t$ and $F_t$.

\subsection{SDN Data Forwarding}
\label{subsec:sdn data forwarding}

We assume  that an SDN data forwarding algorithm is executed whenever a new request $q$ arrives, i.e., the data forwarding is an event-driven (aperiodic) process. In order to handle the continuous stream of requests from network users, which are not a-priori-known, a network system must continuously respond to new requests arriving at any time -- even in the middle of addressing a congestion problem. Our data forwarding algorithm, which is similar to existing baselines~\cite{Bianco:17}, uses weight parameters assigned to network links and computes the weighted shortest path between a pair of switches to determine the route for carrying a data stream of $q$ between the switches. Specifically, we denote by $w(e)$ the weight value of a link $e$. The default weights are one (i.e., $w(e)=1$ for all the links $e$ in $G$). The weights are configurable and can be modified by application layer algorithms.  In Section~\ref{subsubsec:weight}, we discuss how  DICES modifies the weight parameters after detecting congestion  so  that the data forwarding algorithm does not send new requests through the  overutilized links.

\subsection{Dynamic Adaptive Congestion Control}
\label{subsec:dices}

DICES runs in parallel with the SDN data forwarding algorithm described in Section~\ref{subsec:sdn data forwarding}. In contrast to the SDN data forwarding algorithm, DICES is designed to  execute periodically with a time period $\varDelta$. To detect congestion,  DICES has to poll   the network state periodically as the state is always changing due to the unpredictable environment. In addition, DICES has to ensure, when congestion happens, that the subsequent steps for congestion resolution are always deterministically executed. Therefore, we chose to design DICES as a periodic process  instead of an event-driven (aperiodic) one.  The period $\varDelta$ should be chosen such that it is small enough to allow DICES to detect and handle congestion as quickly as possible, and at the same time, large enough for executions of DICES not to cause too much overhead and interfere with other SDN operations, e.g., the execution of the SDN data forwarding algorithm.

Let $\mathbb{T} = [0,T]$ be the time duration during which we observe the network traffic.  We assume the network $G$ is fixed over time, but the network traffic, i.e.,  the set $Q$ of requests and the set $F$ of flows handling $Q$,  vary over time. We denote by $Q_i$ the set of network requests received at the beginning of the time step  $i{\cdot} \varDelta$, and by $F_i$ the set of flows corresponding to $Q_i$. At each time step $i{\cdot} \varDelta$, DICES starts running by executing its ``monitor'' step (Fig.~\ref{fig:overview}). It receives $Q_i$ and $F_i$ and uses these two sets in its subsequent steps, i.e., ``analyze'', ``compute'', and ``apply''. Requests that arrive within the interval of $[i{\cdot}\varDelta, (i{+}1){\cdot}\varDelta)$ or the flows generated within this interval are included in $Q_{i+1}$ and $F_{i+1}$, but not in $Q_i$ and $F_i$.

The ``analyze'' step is in charge of determining whether, or not, the network is congested. In practice, a link $e$ is  considered  congested if it is utilized above a certain threshold (e.g., 80\% of the link bandwidth)~\cite{Lin:16,Akyildiz:14}. We denote by $\mathit{util}(e,F_i)$ the utilization of link $e$ by the flow set $F_i$ and define it as follows: $\mathit{util}(e,F_i) = \mathit{throughput}(e,F_i) / c(e)$. The ``analyze'' step deems $e$ to be congested if $\mathit{util}(e,F_i) > u$, where \hbox{$0 < u \le 1$} is the \emph{utilization threshold}.

%\begin{figure}[t]
%  \centerline{\includegraphics[width=0.8\columnwidth]{figs/controllerBehavior}}
%  \caption{An overview of dynamic adaptive congestion control. \textcolor{red}{This is hard to follow. To discuss.} \textcolor{red}{Something perhaps needs to be said about the features of SDN allowing network reconfiguration to take effect "while" requests are still being processed.}}
%  \label{fig:controller behavior}
%\end{figure}

If the network $G$ is congested as determined by the ``analyze'' step, the ``compute'' step addresses the  congestion problem by performing the following two tasks: First, it resolves the congestion by computing a new set $F_i^a$ of $F_i$ that can handle the requests $Q_i$ without congestion (see the congestion problem definition in Section~\ref{subsec:problem}). If congestion cannot be resolved, it ensures that $F_i^a$ minimizes the maximum link utilization by $Q_i$. Second, it computes a set of weights for network links based on their utilization. These weights are passed to the SDN data forwarding algorithm (Section~\ref{subsec:sdn data forwarding}) so that the algorithm does not send new requests arriving after $i{\cdot} \varDelta$  through the overutilized links.  The ``apply'' step  reconfigures flows and  applies the new weights computed by \hbox{the ``compute'' step.}
%After completing a cycle of Fig.~\ref{fig:overview}, DICES starts again from its ``monitor'' step to receive flows $F_{i+1}$  and requests $Q_{i+1}$ at time step ${(i{+}1)}{\cdot} \varDelta$ since DICES is designed as a periodic process.

%Note that one cycle of the control loop in Fig.~\ref{fig:controller behavior} takes $\delta$ time. Our dynamic adaptive control approach assumes that there are no dramatic changes in network requests within a time $\delta$ period.

In the remainder of this section, we present two algorithms addressing the two tasks of the ``compute'' step: A \emph{search-based congestion control algorithm} for the first task, and a \emph{utilization-aware weight control algorithm} for the second task.

\subsubsection{Search-based Congestion Control Algorithm}
\label{subsubsec:search}

%Due to the nature of dynamic adaptive configuration, our solution considers (1)~its execution time and (2)~the tradeoff between the costs and benefits of configuration. The configuration cost is determined by the number of time-consuming link updates (insertions and deletions) which should be minimized while solving the congestion problem. We cast the problem of network congestion as a multi-objective search-based optimization problem since: (1)~It can be solved by less computationally expensive algorithms than exact optimization algorithms. Note that the congestion problem is a hard problem~\cite{Brandt:16,Amiri:18}. Search-based solutions have been successfully applied in many applications where the goal is to find practically useful (near-)optimal solutions while accounting for scalability considerations~\cite{??}. (2)~A multi-objective search-based approach can provide multiple equally viable solutions to choose from, which are the best tradeoffs found regarding the costs and benefits for dynamic adaptive configuration.

%Depending on applications, system engineers select a solution from the equally viable solutions that, for instance, maximizes the overall network capacity to handle new requests or minimizes the overall network delay to deliver data streams on time.

\begin{sloppypar}
Our search-based congestion control algorithm attempts to resolve an identified congestion, and if the congestion cannot be resolved, the algorithm minimizes the maximum link utilization. Specifically, given a network $G$ congested by the set $F_i$ of flows addressing the set $Q_i$ of requests, our aim is to generate the set $F^a_i$ of flows to resolve or minimize the congestion while addressing the requests in $Q_i$. 
To do so, we minimize the maximum link utilization across all the links in $G$ (objective \emph{O1} or \emph{Utilization}). In addition to minimizing utilization, we aim to optimize two more objectives that are important for quality of service in network systems: We minimize   the number of link updates, i.e., insertions and deletions, required to reconfigure the network flows (objective \emph{O2} or \emph{Cost}) and the overall data transmission delays induced by the new set $F^a_i$ of flows (objective \emph{O3} or \emph{Delay}). By minimizing the cost, we ensure that we   manipulate a small number of   elements at the infrastructure layer and require a small amount of time to apply $F_i^a$. Minimizing the network delay is critical for  emergency  systems to ensure that data streams are transmitted on time.  Note that we have to optimize these three objectives explicitly and simultaneously since optimizing the utilization objective, \emph{O1}, is likely to negatively impact the cost of flow reconfiguration, \emph{O2}, or the overall delay, \emph{O3}. This is because  if the new flow paths of $F_i^a$ are very different from those of $F_i$ or if $F_i^a$ uses longer but less utilized paths than those of $F_i$, the reconfiguration cost and the overall delay may increase. In addition,  the reconfiguration cost, \emph{O2}, and  the overall delay, \emph{O3}, are independent objectives. 
\end{sloppypar}

%For example, suppose the congestion problem of the network $G$ with flows $F_i$ for requests $Q_i$ can be solved by using an alternative set of flows $F_i^a$ for $Q_i$ and $F_i^a$ does not reuse any flows in $F_i$, i.e., $F_i \cap F_i^a = \{\}$. While addressing the congestion, the flow configuration cost, i.e., the number of link updates, from $F_i$ to $F_i^a$ should ideally be small. However, $F_i \cap F_i^a = \{\}$ implies that \emph{all} the flows in $F_i$ should be changed to transform $F_i$ into $F_i^a$. Besides, the overall delay induced by $F_i^a$ can be greater than the delay induced by $F_i$ if the alternative flows $F_i^a$ use longer but less utilized paths than those of $F_i$. Minimizing the flow configuration cost~\emph{O2} may also negatively impact the overall delay~\emph{O3}. For instance, assuming a flow path $(e_1,e_2)$ in Fig.~\ref{fig:eg network} can be reconfigured by either $(e_3,e_4)$ with an overall 500 link delay or $(e_5,e_6,e_8)$ with an overall 75 link delay, the flow configuration cost of $(e_3,e_4)$ is smaller but its overall link delay is greater.

Following standard practice~\cite{Ferrucci:13}, we describe our algorithm by defining the representation, the initial population, the fitness functions, and the computational search algorithm. We then discuss the output flow set $F_i^a$ that we report as the optimal solution to be  used in the ``apply'' step of DICES. 

\noindent\textbf{Representation.} Given a network $G$ and a set $Q$ of requests, a feasible solution is a set $F = \{f_1, f_2, ..., f_l\}$ of flows where for every $f \in F$ we have $f.q \in Q$, for every $q' \in Q$ there is some $f \in F$ such that $f.q = q'$, and $|F| = |Q|$.

%and flows $F_i$, a feasible solution is a set $F_i^a = \{f_1^a, f_2^a, ..., f_l^a\}$ of flows, where $f_k^a$ is an alternative flow corresponding to $f_k \in F_i$ such that both $f_k^a$ and $f_k$ process the same request, i.e., $f_k^a.q = f_k.q$, $f_k.q \in Q_i$, and $|F_i^a| = |F_i|$. Note that the number of alternative flows to a given flow $f_k$ depends on the network $G$. 

\noindent\textbf{Initial population.} Recall that the input to our search algorithm is a set  $F_i$ of flows at time $i{\cdot} \varDelta$ and its corresponding set $Q_i$ of requests.  We create an initial population by randomly modifying individual flows in $F_i$  while ensuring that the generated flow sets are able to handle the requests in $Q_i$. 

\begin{sloppypar}
\noindent\textbf{Fitness.} For the three objectives \emph{O1}, \emph{O2}, and \emph{O3} described above, we formulate three quantitative fitness functions $\mathit{fitUtil}(F_i^a)$, $\mathit{fitCost}(F_i,F_i^a)$, and $\mathit{fitDelay}(F_i^a)$, respectively, where $F_i$ is the set of flows given as input and $F_i^a$ is a candidate flow set generated during the search. 
\end{sloppypar}

The $\mathit{fitUtil}(F_i^a)$ fitness function is defined by equation~(\ref{eq:o1}) as  the maximum link utilization across all the links used in $F_i^a$. Our approach aims to minimize equation~(\ref{eq:o1}). 
\begin{small}
\begin{equation}
\mathit{fitUtil}(F_i^a) = \max_{e \in \mathit{links}(F_i^a)}\mathit{util}(e,F_i^a)
\label{eq:o1}
\end{equation}
\end{small}

The $\mathit{fitCost}(F_i,F_i^a)$ fitness function is defined by equation~(\ref{eq:o2}).  In this paper, we compute the distance between a pair $f$ and $f'$ of flows, denoted by $\mathit{dist}(f,f')$, as the \emph{edit distance} between  the path of $f$  ($f.p$) and  the path of $f'$ ($f'.p$).  Our notion of edit distance is the same as computing the longest common subsequence (LCS) distance of two paths~\cite{Cormen:09} and counts 
 the number of \emph{link insertions} and \emph{link deletions} required  to transform $f.p$  into  $f'.p$. This metric matches our definition of the cost objective described earlier in this section.  Our approach minimizes  equation~(\ref{eq:o2}). 
 \begin{small}
 \begin{equation}
 \mathit{fitCost}(F_i,F_i^a) = 
 \sum\limits_{(f,f') \in F_i {\times} F_i^a: f.q = f'.q} \mathit{dist}(f,f')
 \label{eq:o2}
 \end{equation}
 \end{small}
 
%For example, assuming $f.p = (e_3,e_4)$ and $f'.p = (e_3,e_7,e_8)$ (see Fig.~\ref{fig:eg network}), then $\mathit{dist}(f,f') = 3$. 

%Note that the sum of $dist()$ is equal to $0$ if $F_i^a = F_i$. In this case, obviously, solution $F_i^a$ does not resolve the congestion problem induced by $F_i$. In such case, to make $F_i^a$ infeasible, our fitness computation sets $\mathit{fitCost}()$ value of $F_i^a$ to a large number $\mathit{lrg}$.

%\begin{small}
%\begin{equation}
%\mathit{fitCost}(F_i,F_i^a) = 
%\begin{cases}
%\sum\limits_{\mathclap{\hspace{7em}f \in F_i \wedge f' \in F_i^a \wedge f.q = f'.q}} \mathit{dist}(f,f') &  \mbox{if}~F_i \neq F_i^a\\
%\infty & \mbox{otherwise}\\
%\end{cases}
%\label{eq:o2}
%\end{equation}
%\end{small}

The $\mathit{fitDelay}(F_i^a)$ fitness function is defined by equation~(\ref{eq:o3}) which sums the delay values $l(e)$ of all the links $e$ used in a candidate solution $F_i^a$. Note that the delay objective can be estimated for a flow set $F_i^a$ only if $F_i^a$ does not give rise to congestion, i.e., only when  $\mathit{fitUtil}(F_i^a) \le u$, where $u$ is a utilization threshold. This is because, when a network is congested, actual delay values depend on various factors  such as the underlying network protocol (e.g., TCP or UDP)  that are not studied here. Hence, when $F_i^a$ leads to congestion,  we assign an undefined value (i.e., a large number) to $\mathit{fitDelay}(F_i^a)$. Our approach minimizes equation~(\ref{eq:o3}). 
\begin{small}
\begin{equation}
\mathit{fitDelay}(F_i^a) = 
\begin{cases}
  \underset{e \in \mathit{links}(F_i^a)}{\sum}l(e) & \mbox{if } \mathit{fitUtil}(F_i^a) \le u \\
  \mbox{UNDEF} & \mbox{otherwise}
\end{cases}
\label{eq:o3}
\end{equation}
\end{small}

Recall from Section~\ref{subsec:problem}, that congestion may not be resolved by our approach which is based on reassigning the flows. In this case, the objective $\mathit{fitDelay}()$ is excluded since it is undefined and returns a large number for all the congested solutions. But the search still minimizes  $\mathit{fitUtil}()$ and $\mathit{fitCost}()$  and returns a solution $F^a_i$  that is minimally congested and its implementation incurs minimal cost. 
 
\begin{sloppypar}
\noindent\textbf{Computational search.} We use the Non-dominated Sorting Genetic Algorithm version 2 (NSGAII) algorithm~\cite{Deb:02} to find a near-optimal solution. NSGAII outputs a set (Pareto front) of non-dominated solutions which are equally viable and the best tradeoffs found among the given fitness functions. The dominance relation over solutions is defined as follows~\cite{Knowles:00}: ``A solution $F_i^b$ dominates another solution $F_i^a$ if $F_i^b$ is not worse than $F_i^a$ in all fitness values, and $F_i^b$ is strictly better than $F_i^a$ in at least one fitness.''
\end{sloppypar}

\begin{figure}[t]
\begin{lstlisting}[style=Alg]
Algorithm Search-based congestion control
Input $G$: Network
Input $Q_i$: Set of requests at time $i {\cdot} \varDelta$
Input $F_i$: Set of flows at time $i {\cdot} \varDelta$
Input $u$: Upper threshold of link utilization
Input psize: population ?and? archive size
Input cprob: Crossover probability
Input mprob: Mutation probability
Input neval: Maximum number of evaluations
Output $F_i^b$: Best solution

// initial population
$\mathcal{P}$ $\leftarrow$ $\{F_i\}$ // ?\textcolor{javagreen}{$\mathcal{P}$}? is a set of sets
while $|\mathcal{P}|$ $<$ psize do
?\vrule?  $F_i^a$ $\leftarrow$ mutate($G$, $F_i$)
?\vrule?  $\mathcal{P}$ $\leftarrow$ $\mathcal{P} \cup \{F_i^a\}$
$\mathcal{A}$ $\leftarrow$ $\{\}$ // initial archive
for neval times do
?\vrule?  // fitness evaluation
?\vrule?  for each $F_{ik}^a \in \mathcal{P}$ do
?\vrule?  ?\vrule?  $\mathit{fitUtil}(F_{ik}^a) = \underset{e \in \mathit{links}(F_{ik}^a)}{\max}\mathit{util}(e,F_{ik}^a)$
?\vrule?  ?\vrule?  $\mathit{fitCost}(F_i,F_{ik}^a) = \sum\limits_{(f,f') \in F_i {\times} F_{ik}^a: f.q = f'.q}\mathit{dist}(f,f')$
?\vrule?  ?\vrule?  if $\mathit{fitUtil}(F_{ik}^a) \le u$ then
?\vrule?  ?\vrule?  ?\vrule?  $\mathit{fitDelay}(F_{ik}^a) = \underset{e \in \mathit{links}(F_{ik}^a)}{\sum}l(e)$
?\vrule?  ?\vrule?  else
?\vrule?  ?\vrule?  ?\vrule?  $\mathit{fitDelay}(F_{ik}^a) = \mbox{UNDEF}$
?\vrule?  $\mathcal{P}$ $\leftarrow$ $\mathcal{P} \cup \mathcal{A}$
?\vrule?  $\mathcal{B}$ $\leftarrow$ paretoFront($\mathcal{P}$)
?\vrule?  $\vv{\mathcal{R}}$ $\leftarrow$ sortNonDominatedFronts($\mathcal{P}$)
?\vrule?  $\mathcal{A}$ $\leftarrow$ $\{\}$
?\vrule?  for each front $\mathcal{R}_k$ in $\vv{\mathcal{R}}$ do
?\vrule?  ?\vrule?  assignCrowdingDistance($\mathcal{R}_k$)
?\vrule?  ?\vrule?  // union() returns ?\textcolor{javagreen}{$|\mathcal{A}|~\leq$}? psize
?\vrule?  ?\vrule?  $\mathcal{A}$ $\leftarrow$ union($\mathcal{A}$, $\mathcal{R}_k$, psize)
?\vrule?  ?\vrule?  if $|\mathcal{A}|$ = psize then break
?\vrule?  $\mathcal{P}$ $\leftarrow$ breed($\mathcal{A}$, cprob, mprob)
$F_i^b$ $\leftarrow$ selectOne($\mathcal{B}$)
return $F_i^b$
\end{lstlisting}
\vspace{-1.0em}
\caption{An NSGAII-based congestion control algorithm.}
\Description{An NSGAII-based congestion control algorithm.}
\label{fig:nsgaii}
\vspace{-0.5em}
\end{figure}

Fig.~\ref{fig:nsgaii} presents our NSGAII-based congestion control algorithm. As shown in lines 12--16, we first create an initial population based on the input $F_i$. Lines 19--26 of the algorithm compute the fitness functions. Lines 27--36 describe how NSGAII selects best solutions (lines 27--28), sorts non-dominated fronts (line 29), and assigns crowding distance (line 32) to introduce diversity among non-dominated solutions~\cite{Deb:02}.

\begin{figure}[t]
\begin{lstlisting}[style=Alg]
Algorithm Flow mutation
Input $G$: Network
Input $F_i^a$: Set of flows
Input mprob: Mutation probability
Output $F_i^m$: Set of flows

$F_i^m$ $\leftarrow$ $F_i^a$
for each $f_k \in F_i^a$ do
?\vrule?  if mprob $\ge$ random(0,1) then
?\vrule?  ?\vrule?  $f_k^a$ $\leftarrow$ alternativeFlow($G$, $f_k$)
?\vrule?  ?\vrule?  $F_i^m$ $\leftarrow$ $(F_i^m \setminus \{f_k\}) \cup \{f_k^a\}$
return $F_i^m$
\end{lstlisting}
\vspace{-1.0em}
\caption{A flow mutation algorithm.}
\Description{A flow mutation algorithm.}
\label{fig:mutation}
\vspace{-1.5em}
\end{figure}

As per line 36 of the listing in Fig.~\ref{fig:nsgaii}, the algorithm breeds the next population by using the following genetic operators: (1)~\emph{Selection.} We use the binary tournament selection based on non-domination ranking and crowding distance as typically used by NSGAII~\cite{Deb:02}. (2)~\emph{Crossover.} We use the standard single-point crossover which has been applied in many problems~\cite{Andrade:13,Hemmati:11,Deb:02}. (3)~\emph{Mutation.} We use the mutation algorithm in Fig.~\ref{fig:mutation}. It replaces a randomly selected flow $f_k$ in $F_i^a$ (lines 8--9) with an alternative flow $f_k^a$ for $f_k$ such that $f_k^a.q = f_k.q$ (lines 10--11).

\begingroup

\noindent\textbf{Choosing an optimal solution.} The output of NSGAII is a set of equally viable solutions (line 28 in Fig.~\ref{fig:nsgaii}). But we have to select only one solution  to be used for reconfiguring the network (lines 37--38).  Researchers have proposed various alternatives for selecting an optimal solution among all the solutions on an optimal Pareto front, such as a \emph{knee} solution~\cite{Branke:04} or the \emph{corner} solution~\cite{Panichella:15} for an objective. In our work, we use a knee solution.
\setlength{\columnsep}{1.5em}%
\begin{wrapfigure}[14]{r}{0.4\columnwidth}
  \vspace{-1.8em}
	\centerline{\includegraphics[width=0.38\columnwidth]{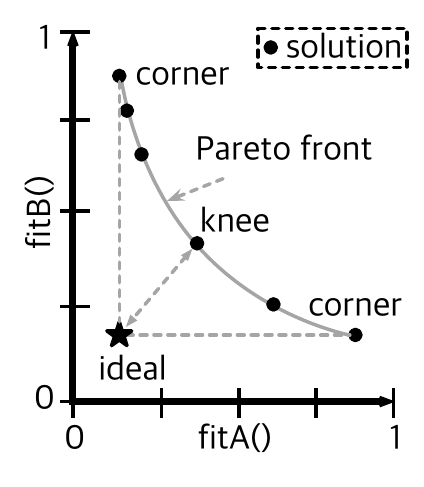}}
  \vspace{-1.5em}
	\caption{The concept of a knee point solution on a two-objective Pareto front.}
  \Description{The concept of a knee point solution on a two-objective Pareto front.}
%  \vspace{-1.0em}
	\label{fig:knee}
\end{wrapfigure}
Fig.~\ref{fig:knee} illustrates the knee point on a two-objective Pareto front. Specifically, a knee point is the closest point on a Pareto front to the \emph{ideal point}, a hypothetical solution with the best values for all the objectives.  A knee solution is often preferred in SBSE studies~\cite{Chen:18,Branke:04} because a small improvement in one objective by selecting other solutions on the front would lead to a large deterioration in at least one other objective~\cite{Branke:04}.  In  our work,  we choose a knee solution  because we do not want the selected flow set to be biased toward any objective; instead, we prefer a flow set that is equally optimized for all the objectives. 

\endgroup

%For example, a solution that marginally improves $\mathit{fitUtil}()$ may lead to significant deterioration of $\mathit{fitCost}()$ or $\mathit{fitDelay}()$. Such biased solutions are unacceptable because, for example, severe deterioration of $\mathit{fitDelay}()$ would likely lead to other practical problems related to delays.

\subsubsection{Utilization-aware Weight Control}
\label{subsubsec:weight}

%\begin{figure}[t]
%  \centerline{\includegraphics[width=0.7\columnwidth]{figs/egWeight}}
%  \caption{An example of a congested network, i.e., $e_c$ is over-utilized $\mathit{util}(e_c,F_i) > u$, and the three requests $q_a$, $q_b$, and $q_c$ are arriving after $i {\cdot} \varDelta$ time. Note that all the links have the same delay.} \textcolor{red}{How does this example show that $e_c$ is overutilised? Either the figure or the caption needs to be improved.}
%  \label{fig:eg weight}
%\end{figure}

As described in Section~\ref{subsec:sdn data forwarding}, the SDN data forwarding algorithm constructs a flow that always routes a data stream along the weighted shortest path. A network system that uses static link weights, e.g., $w(e) = 1$ for all the links $e$ in $G$, may remain congested even after applying our search-based solution in Section~\ref{subsubsec:search}. This occurs when a highly utilized link keeps being used  by the SDN data forwarding algorithm to carry new data requests  because the link is located on a weighted shortest  path. Note that, in general, links located on the shortest paths are more likely to be highly utilized or congested when we use the SDN data forwarding algorithm with fixed weight values. To avoid this problem, we update link weights in a way that forces the SDN data forwarding algorithm to prioritize less utilized links over highly utilized ones. 
 
 %Fig.~\ref{fig:eg weight} shows an example (partial) network with over-utilized link $e_c$, i.e., $\mathit{util}(e_c,F_i) > u$, and the requests $q_a$, $q_b$, and $q_c$ arriving after the monitoring time step $i {\cdot} \varDelta$. Assuming the link delay values of all the links in Fig.~\ref{fig:eg weight} are the same and the weighted shortest paths for $q_a$, $q_b$, and $q_c$ pass through the link $e_c$, this link can remain congested while there are alternative (less utilized) paths, e.g., $(e_k,e_l,e_m)$. Even though such congestion problem can be resolved by our search-based congestion control algorithm described in Section~\ref{subsubsec:search} when DICES detects the congestion, e.g., after $(i+1){\cdot}\varDelta$, the congestion will deteriorate the network performance. To avoid such possible congestion problem, we develop a utilization-aware weight control algorithm which is executed immediately after executing the search-based congestion control algorithm.

\begin{figure}[t]
\begin{lstlisting}[style=Alg]
Algorithm Link weight adjustment
Input $F_i^b$: Solution of the algorithm ?in? Fig.?~\ref{fig:nsgaii}?
Input $\vv{w}_i$: Vector of link weights at time $i {\cdot} \varDelta$
Input $u$: Upper threshold of link utilization
Output $\vv{w}^o$: Vector of adjusted link weights

$\vv{w}^o$ $\leftarrow$ $\vv{w}_i$
for each $e \in \mathit{links}(F_i^b)$ do
?\vrule?  $w(e)$ = $l(e) {\cdot} u / (u - \mathit{util}(e,F_i^b))$
?\vrule?  $\vv{w}^o$ $\leftarrow$ replaceWeight($\vv{w}^o$,$e$,$w(e)$)
return $\vv{w}^o$
\end{lstlisting}
\vspace{-0.8em}
\caption{A link weight adjustment algorithm.}
\Description{A utilization-aware link weight adjustment algorithm.}
\label{fig:weight}
\vspace{-1.5em}
\end{figure}

Fig.~\ref{fig:weight} describes our utilization-aware link weight adjustment algorithm. The algorithm modifies the link weights based on the statement on line~9. Specifically, it adjusts the weights such that $w(e)$ of a link $e$ is proportional  to link delay $l(e)$, but is inversely proportional to the remaining (available) bandwidth of the link $e$, i.e., $u / (u - \mathit{util}(e, F_i^b))$. Note that the bandwidth is computed after applying the new optimized flows $F_i^b$ generated by the search-based algorithm in Fig.~\ref{fig:nsgaii}. The weight computation thus assigns a large value to a highly utilized but lower-speed link (i.e., a link with large delay). The SDN data forwarding algorithm (see Section~\ref{subsec:sdn data forwarding}) then selects less utilized and higher-speed links when it creates flows to address new requests arriving after \hbox{the weight adjustment, i.e., after $i {\cdot} \varDelta$.}

%Fig.~\ref{fig:weight} describes the utilization-aware weight adjustment algorithm. The algorithm modifies the standard link weight heuristic which makes the weight of a link inversely proportional to its bandwidth~\cite{??}, i.e., \emph{link weight} = \emph{reference bandwidth} / \emph{link bandwidth}, e.g., \emph{reference bandwidth} = 100 Mbps. Specifically, the algorithm adjusts the weight value of a link $\mathit{weight}(e)$ by equation~(\ref{eq:weight}) which is based on the \emph{link delay} $l(e)$, of \emph{utilization threshold} $u$, and \emph{current utilization} $\mathit{util}(e, \vv{b})$ (see line 10--11). The rationale behind the weight computation is that we assign a large value to a highly utilized and low-speed link. The ``process requests'' step in Fig.~\ref{fig:controller behavior} will then select less utilized and high-speed links for handling new requests.

%\begin{small}
%\begin{equation}
%w(e) = l(e) \cdot u / (u - \mathit{util}(e, F_i^b))
%\label{eq:weight}
%\end{equation}
%\end{small}

%One can modify the flows by modifying the weight. 
%Our approach assigns a weight value to each link based on link utilization (described in Section~\ref{subsubsec:weight}). The weighted shortest paths from $q.s$ to $q.t$ are then computed using the weight values. A weighted shortest path algorithm is commonly used in network systems to construct new flows since it is solved in polynomial time~\cite{??}.
% !TEX root =  paper.tex

\section{Empirical Evaluation}
\label{sec:exp}

In this section, we present an evaluation of DICES. Our full evaluation package is available online~\cite{Artifacts}.

\subsection{Research Questions (RQs)}
\label{subsec:rq}

\noindent\textbf{RQ1 (efficiency and effectiveness)}: \emph{Can our approach resolve congestion caused by changes in network requests over time?}   In RQ1, we examine the efficiency and effectiveness of DICES by investigating whether it is able to detect congestion as we increase network requests, and whether it can compute and apply an adequate reconfiguration in a timely manner.

\noindent\textbf{RQ2 (scalability)}: \emph{Can our approach resolve congestion promptly for large-scale networks?}  In RQ2, we investigate the scalability of DICES by studying the relation between its execution time and the network size and number of requests. 

%To answer RQ2, we create a number of different networks with different sizes and different requests and investigate the time required by DICES to resolve congestions in the networks as a function of these parameters. We further measure the execution time of DICES when it addresses a congestion problem occurred in the EMS network.

\noindent\textbf{RQ3 (comparison with baselines)}: \emph{How does our approach perform compared with baseline approaches?} With  RQ3, we investigate whether our approach can outperform two existing packet forwarding algorithms: a reactive forwarding algorithm (RFWD)~\cite{Bianco:17} and an open shortest path first algorithm (OSPF)~\cite{Coltun:08}. RFWD and OSPF, discussed in Section~\ref{subsubsec:exp}, are commonly used for optimal data forwarding and congestion avoidance, respectively, and as baselines in recent SDN research strands~\cite{Poularakis:19,Amin:18,Bianco:17,Rego:17,Caria:15,Bianco:15,Agarwal:13}.
%\cite{Bianco:17,Bianco:15,Agarwal:13,Ericsson:02,Fortz:00}

\subsection{Simulation Platform}
\label{subsec:imp}
We implemented DICES as an application for an SDN testbed at SES. Specifically, we use  an open-source SDN control platform known as  ONOS (Open Network Operating System)~\cite{Berde:14}.  ONOS has been used extensively in research and practice~\cite{Bianco:17,Bakhshi:17}, in particular for large-scale network systems. To simulate networks, we use Mininet~\cite{Lantz:10} and D-ITG~\cite{Botta:12}. Mininet is a network emulator that creates a virtual network, running real SDN-switch and application programs, on a single machine to ease prototyping and testing. D-ITG (Distributed Internet Traffic Generator) is a traffic generation and monitoring tool that supports various network protocols and traffic distributions for replicating realistic network traffic.
We ran all our experiments on a computer equipped with an Intel i7 CPU with 8GB of memory.

%We used this platform to simulate a number of synthetic network systems as well as an EMS profile specified by SES. In the description of our simulation platform and our evaluation, we omit several network-specific details of the control and infrastructure layers that are not relevant for our purpose, e.g., packet forwarding in a switch, and evaluate our work solely based on the impact observed at the application layer. 

\subsection{Study Subjects}
\label{subsec:subject}

We use two types of study subjects: (1)~some synthetic networks, and (2)~EMS -- a large-scale industrial system under study by SES (see Section~\ref{sec:motivation}). The synthetic networks are used to evaluate efficiency, effectiveness and scalability since, in these networks, we can freely change the size and traffic, while EMS is used to evaluate the execution time of DICES and to compare it with baselines in a realistic setting.

Our synthetic networks are characterized by two parameters: the number of network switches and the number of network requests. We assume  complete graph topologies, i.e.,  all the switches are connected to one another using links with 100Mbps bandwidth and 25ms delay. Hence, the number of links is not an open parameter for our synthetic networks. This choice was made to reduce unnecessary complexity in our analysis. The network bandwidth and delay values and network-request profiles were suggested by SES based on the typical characteristics of  terrestrial links and their data streams. We use UDP and TCP -- typical protocols for Internet applications -- for transmitting data. Note that graph topologies, which contain only one path from a sender to receiver, e.g., star and line topologies, are not considered in our experiments. Congestion for such topologies can only be handled by decreasing the network traffic. In contrast, DICES requires multiple flow paths from a sender to a receiver for rerouting. A partially complete graph topology is considered in the EMS network (Fig.~\ref{fig:ems topo}).
 
As discussed in Section~\ref{sec:motivation}, the EMS network contains seven SDN switches, four site types (denoted RM, MC, CS, and GS in Fig.~\ref{fig:ems topo}). The network further contains both terrestrial and satellite links and is connected to \emph{external (legacy) networks (EN)}. The characteristics of the terrestrial and satellite links are as follows: 100 Mbps bandwidth, 25ms delay for terrestrial links, and 10 Mbps bandwidth, 275ms delay for satellite links.

\subsection{Evaluation Metrics}
\label{subsec:metrics}

To answer the RQs, we measure the following network performance metrics: \emph{link utilization}, \emph{packet loss}, and \emph{packet delay}. In addition, we measure the execution time of DICES. The link utilization metric is the maximum link utilization across all the links in a network since a single overutilized link can create congestion. Specifically, given a set $F$ of flows, we compute this metric as the maximum of $\mathit{util}(e,F)$ for every link $e$ (see Section~\ref{subsec:dices} for the definition of $\mathit{util}(e,F)$).

%	by equation~(\ref{eq:max utilization}) the maximum link utilization of $F_i$, $\mathit{maxUtil(F_i)}$.
%\begin{equation}
%\mathit{maxUtil(F_i)} = \max_{e \in \mathit{links}(F_i)}\mathit{util}(e,F_i)
%\label{eq:max utilization}
%\end{equation}

To measure the packet loss and delay metrics, we rely on an existing network monitoring tool, D-ITG, described in Section~\ref{subsec:imp}. Briefly, the packet loss metric for a flow measures the number of packets dropped associated with the flow over a time period, e.g., time interval $\varDelta$. The delay metric for a flow measures each individual packet delivery time from the sender to \hbox{the receiver of the flow.}

Due to random variation in DICES and the traffic generator (D-ITG), we repeat our experiments 50 times. To statistically compare our results, we use  Mann-Whitney \hbox{U-test}~\cite{Mann:47} which determines whether two independent samples are likely or not to belong to the same distribution. We set the level of significance, $\alpha$, to 0.05.

% and Vargha and Delaney's \^{A}\textsubscript{12}~\cite{??}.
%Mann-Whitney U-test is a nonparametric test
%Vargha and Delaney's \^{A}\textsubscript{12} is used to measure the effect of different adaptation strategies for configuration on performance metrics. Note that, two samples of metric values are considered to be equivalent when the value of \^{A}\textsubscript{12} is 0.5.

To determine correlations between the execution time of DICES and the network parameters in our study, i.e., network size and the number of requests, we use regression analysis~\cite{Mosteller:77}. We use $R^2$~\cite{Wright:21} to evaluate the goodness of fit for our regression analysis, providing the proportion of the variance in execution time that can be explained by a regression model.

%F-test~\cite{??}, residual standard error $S$~\cite{??}, and
%To evaluate regression models, we use R-squared value $R^2$~\cite{??}. F-test is used to determine whether independent variables of a regression model are statistically significant to explain changes of the execution time of DICES. We set the level of significance, $\alpha$, to 0.05. To measure the spread around the regression line, we use the residual standard error $S$, which is the standard deviation of the differences between observed execution times and their corresponding predictions from the regression model. $R^2$ is used to compute the proportion of the variance in execution time which is explained by a regression model.

\subsection{Experimental Setup}
\label{subsubsec:exp}

\textbf{EXP1.} To answer RQ1, we create a synthetic network with five switches and generate two network requests between a fixed pair of switches every 10s. Each request transmits 30Mbps of data. Hence, the volume of data that the network transmits increases over time, i.e., 60Mbps initially, 120Mbps after 10s, 180Mbps after the next 10s, and so on.

\textbf{EXP2.}  To answer RQ2,  we  perform two analyses: (1)~To study the correlation between the execution time of DICES and the network size,  we create ten synthetic networks with 5, 10, \ldots, 50 SDN switches, and for each network, we generate five requests simultaneously to transmit a total of 150Mbps data. (2)~To study the correlation between the execution time of DICES and the number of requests, we use a network with five switches, and perform ten different experiments by issuing  5, 10, \ldots, 50 requests simultaneously to transmit, each time, a total of 150Mbps. We also compute the execution time of DICES over the EMS network.

 %To evaluate the adaptability of our approach in detail, we create two variants of our approach \emph{SW} named \emph{NW} and \emph{SN}. The \emph{NW} and \emph{SN} variants are created by excluding either of the ``solve current congestion'' step or the ``solve future congestion'' step, respectively, in Fig.~\ref{fig:controller behavior} from our approach \emph{SW}.

\begin{table}
	\caption{A traffic profile for a disaster situation.}
	\Description{A traffic profile for a disaster situation.}
	\vspace{-1.2em}
	\label{tbl:traffic profile}
	\fontsize{8}{8}\selectfont
	\def\arraystretch{0.8}%  1 is the default, change whatever you need
	\centering
	\begin{tabularx}{1.0\columnwidth}{lllclc}
		\toprule
		\multicolumn{2}{c}{EMS entity} &
		\multicolumn{4}{c}{Request characteristics}\\
		\cmidrule(lr){1-2} \cmidrule(lr){3-6}
		Sender & Receiver & Type & Protocol & Throughput & \# requests\\ 
		\midrule
		RM & MC & Sensor & TCP & 100 Kbps & 5\\
		CS & MC & Audio & UDP & 64 Kbps & 4\\
		CS & MC & Video & UDP & 10 Mbps & 2\\
		MC & CS & Audio & UDP & 64 Kbps & 4\\
		MC & CS & Video & UDP & 10 Mbps & 2\\
		MC & CS & Map & TCP & 30 Mbps & 1\\
		EN\textsuperscript{N} & EN\textsuperscript{D} & External & UDP & 20 Mbps & 5\\
		EN\textsuperscript{D} & EN\textsuperscript{N} & External & UDP & 20 Mbps & 5\\
		\bottomrule
	\end{tabularx}

	\justify The EMS entities are: RM (\emph{remote monitoring site}), 
	MC (\emph{emergency monitoring center}), CS (\emph{mobile communication facility site}), EN\textsuperscript{N} (\emph{external networks} in normal areas), and EN\textsuperscript{D} (\emph{external network} in a disaster area). The disaster area (D) is assumed to be close to s1 in the network of Fig.~\ref{fig:ems topo}. Other than s1, all switches in this network are in normal areas (N).
	
\vspace{-1.8em}
\end{table}

\textbf{EXP3.} To answer RQ3, we use the data traffic profile shown in Table~\ref{tbl:traffic profile}  and defined by SES for the EMS network of Fig.~\ref{fig:ems topo}. The profile characterizes anticipated traffic at the time of a disaster. It includes 28 requests which transmit sensor, audio, video, map, and external data where TCP is used for sensor and map data and UDP  for the rest. In this experiment, we assume a disaster occurs in the area near the s1 switch in Fig.~\ref{fig:ems topo}. Thus, the \emph{mobile communication facility site} is connected to s1.

\begin{sloppypar}
We compare DICES with a reactive forwarding algorithm (RFWD)~\cite{Bianco:17} and an open shortest path first forwarding algorithm (OSPF)~\cite{Coltun:08}. RFWD, which is the only predefined reactive data forwarding  application in ONOS,  routes  requests through the shortest paths between the requests' ending points. It is the same as the SDN data forwarding algorithm in Section~\ref{subsec:sdn data forwarding} when link weights are all equal to one and are fixed all the time. OSPF is commonly used as a baseline for SDN-based network solutions~\cite{Poularakis:19,Amin:18,Rego:17,Caria:15,Agarwal:13}. It is a prevalent protocol in legacy (non-SDN) networks while SDN is still an emerging area in both research and practice. OSPF computes weighted shortest paths to route network requests, but  it does not provide the flexibility to update the link weights dynamically.  We compare DICES with OSPF  when the link weights for OSPF  are inversely proportional to the bandwidths of the links. This is a typical use case of OSPF and  can reduce the possibility of  congestion since high-bandwidth links tend to be more used to carry data. To our knowledge, DICES is the first SDN application available online which addresses a congestion problem while accounting for minimizing multiple objectives: transmission delays and \hbox{reconfiguration costs~\cite{Artifacts}.}
\end{sloppypar}

\subsection{Parameter Tuning and Setting}
\label{subsec:param}

%\begin{figure}[t]
%	\centerline{\includegraphics[width=0.9\columnwidth]{figs/comNumEvals}}
%	\caption{Comparing Euclidean distance values between knee solutions and the reference solution by varying the number of fitness evaluations.}
%	\label{fig:compNumEvals}
%\end{figure}

We set $\varDelta$, i.e., the time period for executing DICES, to 1s since this is the minimum monitoring time period allowed by ONOS. Following the guidelines in the literature~\cite{Arcuri:11}, we set the NSGAII parameters as follows:  the population size~=~100, the crossover probability~=~0.8, and the mutation probability~=~$1/|F_i|$. We set the utilization threshold to 0.8, as instructed by SES.  We set the total number of fitness evaluations to 10,000 because our initial experiments, performed on EMS, showed that, after 10,000 fitness evaluations, there is no notable improvement \hbox{in the optimal solution.}

%We thus set the maximum number of fitness evaluations to 10,000 for our experiments. we ran our NSGAII-based search algorithm for 50 times with 50,000 fitness evaluations. Our results 

%We then determined the minimum and maximum fitness values of each objective from the total of 50,000$\times$50 evaluations. Fig.~\ref{fig:compNumEvals} compares the Euclidean distance values between normalized fitness values of knee solutions, e.g., ($\mathit{fitUtil}()$ - min. utilization) / (max. utilization - min. utilization), and zero fitness values for the three objectives of the reference (ideal) solution. As shown in Fig.~\ref{fig:compNumEvals}, 

%Note that the above parameters can be further optimized. The current setting, however, is clearly adequate for our experiments.
% and other settings do not change the answers to the RQs.

\subsection{Experiment Results}
\label{subsec:results}

\begin{figure*}[t]
	\centerline{\includegraphics[width=0.85\textwidth]{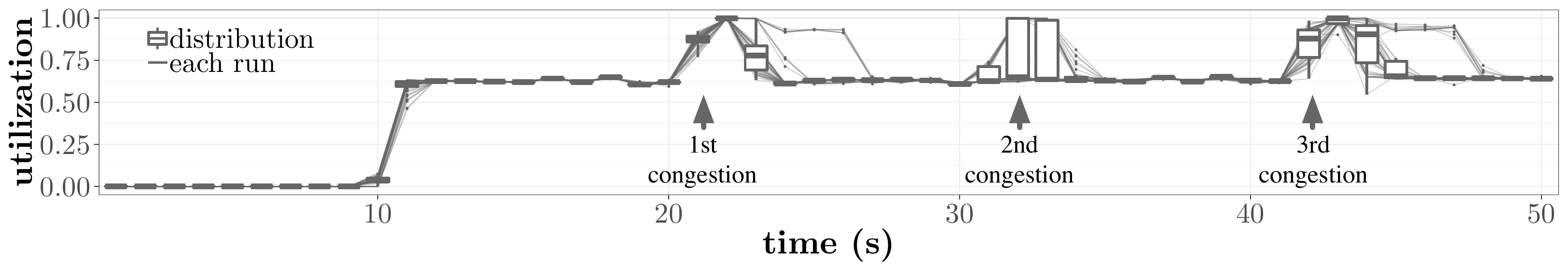}}
	\vspace{-1.2em}
	\caption{Network utilization values over time when DICES is used to resolve congestion for a synthetic network with five switches and when new UDP requests  arrive around 10s, 20s, 30s and 40s. The boxplots (25\%-50\%-75\%) show  network utilization values obtained based on 50 executions of DICES.}
	\Description{Network utilization values over time when DICES is used to resolve congestion for a synthetic network with five switches and when new UDP requests  arrive around 10s, 20s, 30s and 40s. The boxplots (25\%-50\%-75\%) show  network utilization values obtained based on 50 executions of DICES.}
	\label{fig:rq1}
  \vspace{-1.5em}
\end{figure*}

\textbf{RQ1.} Fig.~\ref{fig:rq1} shows the network utilization over time when DICES is used to resolve congestion for the synthetic network described in {\bf EXP1} (see Section~\ref{subsubsec:exp}). As shown in the figure, network requests cause congestion after 20s, 30s and 40s. Note that the requests arriving around 10s do not lead to any congestion and they can be handled by the network. DICES is able to resolve every congestion since utilization always comes back down to around 65\% after the sudden increase caused by each congestion. DICES is further able to resolve congestion in a timely manner. Specifically, it takes  DICES, on average, 439ms to execute all the four steps in its control loop. We note that it takes, on average, 2.68s for the network utilization to settle back to a desired value below the utilization threshold (i.e., 0.8) after congestion. This is due to the additional internal processing time required by ONOS to monitor the network and reconfigure the SDN control and infrastructure layers.

%, which is significantly higher than the execution time of DICES (439ms on average)

As suggested by its low utilization average in Fig.~\ref{fig:rq1}, the second occurrence of congestion around 30s is observed only in  17 out of 50 runs of DICES. More precisely, in the other 33 runs, the link weight adjustment performed by DICES at 20s is able to handle the requests at 30s without leading to any congestion. This is because the link weights adjusted by DICES at 20s can sometimes, due to luck,  help the SDN data forwarding algorithm (Section~\ref{subsec:sdn data forwarding}) handle the requests arriving at 30s using less utilized links, hence preempting congestion.  Note that Fig.~\ref{fig:rq1} shows the results for UDP packet transmission. The results for TCP  packet transmission are  consistent with those in Fig.~\ref{fig:rq1} and not shown due to space.

\begin{mdframed}[style=RQFrame]
	\emph{The answer to {\bf RQ1} is that}  DICES efficiently and effectively resolves congestion. In particular, experiments performed on a realistic network transmitting large and increasing volumes of data over time show that DICES is able to maintain, most of the time, the network utilization at  65\%, which is well below the utilization threshold of 80\%. Further, DICES takes, on average, 439ms to execute and resolve congestion.
\end{mdframed}

\begin{figure}[t]
	\centering
	\subfigure[Links]{%
		\resizebox{0.9\columnwidth}{!}{%
			\scalebox{1}{%
				\includegraphics[scale=1]{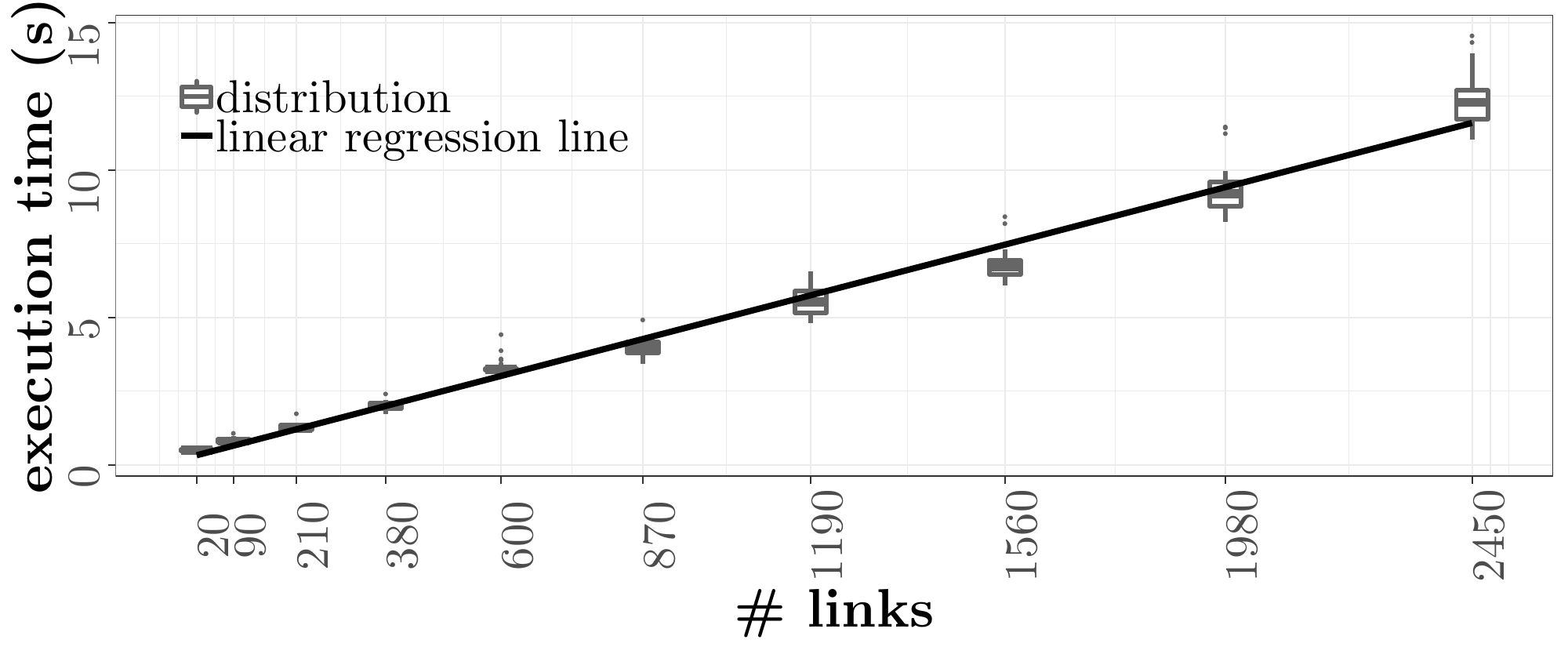}}
			\label{fig:scale topo}
	}}
	\subfigure[Requests]{%
		\resizebox{0.9\columnwidth}{!}{%
%		\vspace{-1.0em}
			\scalebox{1}{%
				\includegraphics[scale=1]{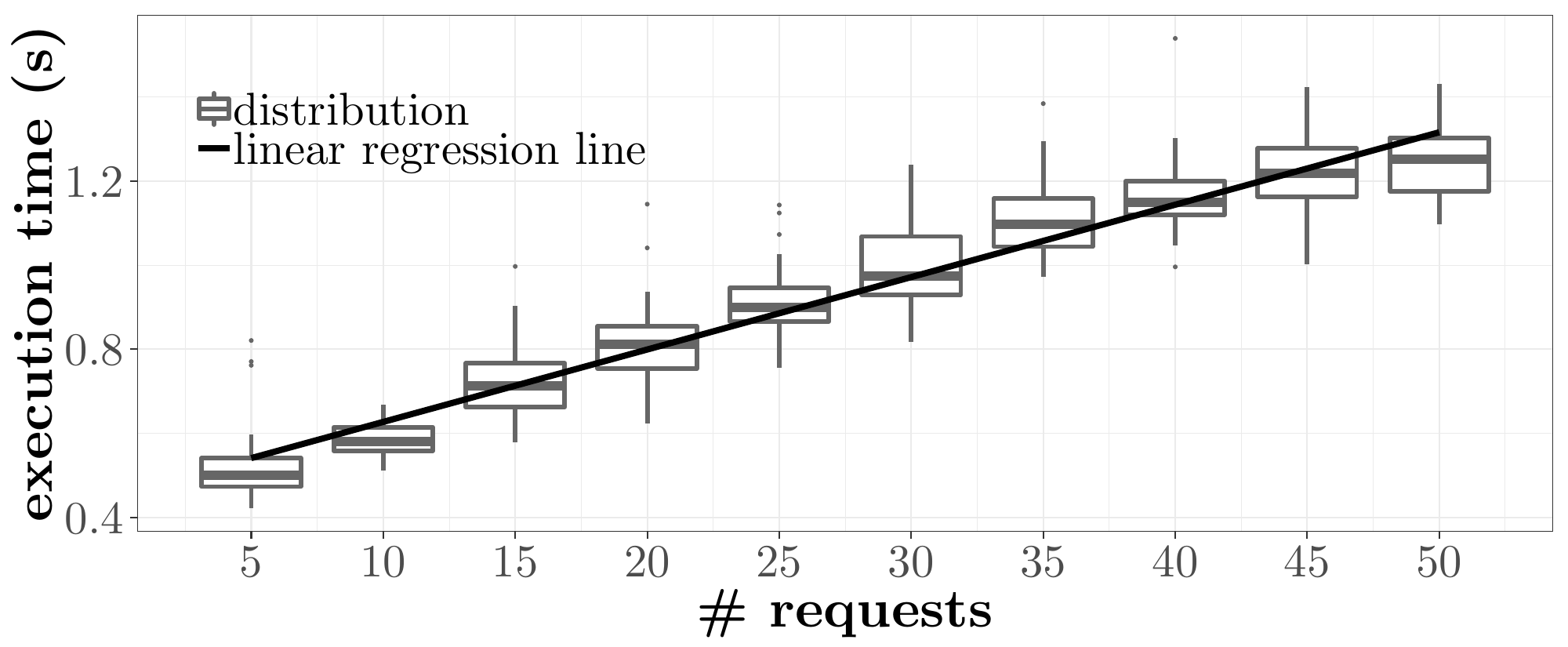}}
			\label{fig:scale req}
	}}
	\vspace{-1.3em}
	\caption{Graphs showing DICES execution time versus (a)~the number of network links, and (b)~the number of network requests together with regression lines showing linear correlation between DICES execution time and (a)~the number of  links ($\mathit{exec. time} {=} 2.391e\text{-}01 {+} 4.635e\text{-}03 {\times} |\mathit{links}|$), and (b)~the number of requests ($\mathit{exec. time} {=} 0.455 {+} 0.017 {\times} |\mathit{requests}|$).}
	\Description{Graphs showing DICES execution time versus (a)~the number of network links, and (b)~the number of network requests together with regression lines showing linear correlation between DICES execution time and (a)~the number of  links ($\mathit{exec. time} {=} 2.391e\text{-}01 {+} 4.635e\text{-}03 {\times} |\mathit{links}|$), and (b)~the number of requests ($\mathit{exec. time} {=} 0.455 {+} 0.017 {\times} |\mathit{requests}|$).}
	\label{fig:rq2}
	\vspace{-1.0em}
\end{figure}

\begin{table*}[t]
	\caption{Comparison of the average delay and packet loss of DICES against RFWD and OSPF based on 50 runs of each algorithm.}
	\vspace{-1.2em}
	\label{tbl:comp alg}
	\fontsize{8}{8}\selectfont
	\centering
	{\renewcommand{\arraystretch}{0.8} %<- modify value to suit your needs
	\begin{tabularx}{1.0\textwidth}{l l @{}Y@{}Y@{}Y@{}Y@{}Y@{}Y}
		\toprule
		& &
		\multicolumn{2}{c}{RFWD} &
		\multicolumn{2}{c}{OSPF} &
		\multicolumn{2}{c}{DICES}\\
		\cmidrule(lr){3-4} \cmidrule(lr){5-6} \cmidrule(lr){7-8}
		Receiver & Type &
		Delay (s) [$p$] & Packet loss (\%) [$p$] & Delay (s) [$p$] & Packet loss (\%) [$p$] & Delay (s) & Packet loss (\%)\\ 
		\midrule
		\multirow{3}{*}{\shortstack[l]{Emergency monitoring\\ center}}
		& Sensor & 0.10 [0.00] & 0.00 [1.00] & 0.09 [0.00] & 0.00 [1.00] & 0.13 & 0.00\\
		& Audio & 0.03 [0.00] & 0.17 [0.00] & 0.03 [0.49] & 0.20 [0.02] & 0.07 & 0.23\\ 
		& Video & 0.03 [0.00] & 0.02 [0.03] & 0.03 [0.30] & 0.02 [0.02] & 0.04 & 0.08\\
		\cmidrule{1-8}
		\multirow{3}{*}{\shortstack[l]{Mobile communication\\ facility site}}
		& Audio & 0.03 [0.00] & 0.17 [0.01] & 0.06 [0.02] & 0.38 [0.00] &  0.08 & 0.18\\
		& Video & 0.03 [0.00] & 0.06 [0.53] & 0.06 [0.00] & 0.32 [0.00] &  0.04 & 0.27\\ 
		& Map & 0.23 [0.00] & 0.00 [1.00] & \cellcolor{light-gray}{\textbf{1.13 [0.00]}} & 0.00 [1.00] & \cellcolor{light-gray}{\textbf{0.10}} & 0.00\\
		\cmidrule{1-8}
		External network
		& External & 0.35 [0.00] & \cellcolor{light-gray}{\textbf{29.55 [0.00]}} & 0.05 [0.00] & 0.14 [0.00] & 0.06 & \cellcolor{light-gray}{\textbf{0.02}}\\
		\midrule\midrule
		Weighted average
		& Overall & 0.29 & 21.81 & 0.17 & 0.13 & 0.06 & 0.04\\
		\bottomrule
	\end{tabularx}
}
\vspace{-1.8em}
\end{table*}

\textbf{RQ2.} Fig.~\ref{fig:rq2} reports the results obtained by {\bf EXP2}. Specifically, Fig.~\ref{fig:scale topo} shows the relation between the execution time of DICES versus network size specified as the number of links (i.e., the first study of  {\bf EXP2}), and  Fig.~\ref{fig:scale req} shows the relation between the execution time of DICES versus the number of requests (i.e., the second study of  {\bf EXP2}). Note that the $x$-axis of Fig.~\ref{fig:scale topo} shows  the number of links instead of the number of switches since DICES mainly manipulates links, and its execution time depends on the number of links and not the number of switches (see the algorithm in Fig.~\ref{fig:nsgaii}). The linear regression lines in both Fig.~\ref{fig:scale topo} and Fig.~\ref{fig:scale req} fit well the actual execution time of DICES with high goodness of fit (i.e.,  $R^2{=}0.98$ for Fig.~\ref{fig:scale topo} and $R^2{=}0.89$ for Fig.~\ref{fig:scale req}). Hence, the execution time of DICES is linear both  in the number of links and in the number of requests. Therefore, we expect DICES to scale well as the numbers of network links  and requests increase.  Finally, for our industrial EMS, which contains seven switches and  30 links, and has to handle 28 requests (see Section~\ref{subsubsec:exp}), DICES, on average, takes 1.74s to resolve congestion. This shows that DICES is able to scale to real-world systems and can resolve congestion caused by high network demands due to an emergency.

We note that our analysis above is concerned with the relation between the execution time of DICES and the number of requests, rather than the data size of network requests. Since DICES resolves congestion by  rerouting requests and never modifies the data size of a request, the execution time of DICES is not impacted by data size. We have confirmed this through experiments that we cannot report due to space.

\begin{mdframed}[style=RQFrame]
	\emph{The answer to {\bf RQ2} is that} the execution time of DICES is linear in the network size and in the number of requests. Further, DICES scales to real-world systems: it takes an average of 1.74s to resolve congestion caused by an emergency situation in our industrial case study.
\end{mdframed}

\textbf{RQ3.} Table~\ref{tbl:comp alg} shows the average delay and packet loss values for EMS when one uses DICES, RFWD and OSPF for handling the requests described in Table~\ref{tbl:traffic profile}. As discussed in Section~\ref{subsubsec:exp}, each experiment was repeated 50 times. The table statistically compares DICES against RFWD and OSPF with respect to delay and packet loss by reporting $p$-values.

In the table, we have highlighted in gray two specific delay values of DICES and OSPF, and two specific packet loss values of DICES and RFWD. These two pairs are particularly interesting because they show significant differences between DICES and OSPF in terms of delay and between  DICES and RFWD in terms of packet loss ($p$-value $<$ 0.05 in both cases). These differences are significant, not just statistically but also practically.  Specifically, the difference in delay values shows that the EMS network with OSPF transmits map data with a 1.13s delay on average.  In contrast, with DICES, the network transmits the map data with a 0.1s delay on average.  In other words, for map-data transmission, the network with DICES is 11 times faster than the network with OSPF. 
The difference in packet loss values shows that the EMS network with RFWD drops, on average, 29.55\% packets while exchanging data between  external  networks (see Table.~\ref{tbl:traffic profile}). When DICES is used,  the network drops only 0.02\% of those packets on average. This shows that DICES is considerably more effective than RFWD in transmitting external data through the EMS network during an emergency situation.
We note that the behavior of DICES is independent of traffic types -- sensor, audio, video, map, and external. As shown in Table~\ref{tbl:comp alg}, DICES maintains a practically acceptable level of delay and packet loss for all types of traffic even when the EMS network is congested. In contrast, the two baselines, i.e., RFWD and OSPF, fail to maintain the level of delay and packet loss at an acceptable level.

%between  EN\textsuperscript{N} and EN\textsuperscript{D}

To compare the overall performance of RFWD and OSPF with DICES, we compute weighted averages of delay and packet loss. Specifically, the weighted average delay (resp., packet loss) for each algorithm is computed by multiplying the average delay (resp., packet loss) of that algorithm for each network request type with the total throughput of that request type (see the request types and throughputs in Table.~\ref{tbl:traffic profile}). The  weighted averages, given in the last row of Table~\ref{tbl:comp alg}, show that  DICES yields lower overall delay and packet loss  compared to both  RFWD and OSPF. 
That is, the overall delay of DICES is almost five and three times better than the overall delays of RFWD and OSPF, respectively. Further, DICES loses almost 99\% and 70\% less packets compared to RFWD and OSPF, respectively.

%on average,  28 links on
We note that the improvements brought about by DICES come at the expense of reconfiguring some flows to resolve congestion, while RFWD and OSPF do not require any reconfiguration. SES found the minimized reconfiguration cost of DICES to be an acceptable tradeoff for the substantial benefits of the approach over RFWD and OSPF in terms of delay and packet loss.

\begin{mdframed}[style=RQFrame]
	\emph{The answer to {\bf RQ3} is that} DICES significantly outperforms the baseline algorithms:  RFWD and OSPF. Specifically, results obtained by simulating emergency traffics over  the EMS network show that the overall network delay of DICES is almost five and three times better than those  of RFWD and OSPF, respectively. Further, DICES loses almost 99\% and 70\% less packets compared to RFWD and OSPF, respectively.
\end{mdframed}

%\subsection{Threats to Validity}
%\label{subsec:threats}

%The most related validity concerns to our work are internal and external validity, as we discuss below.

%\textbf{Conclusion validity}: To mitigate the threats to conclusion validity, we follow a standard guideline in SBSE~\cite{??}. We account for random variations in our experiment results by executing DICES and the other existing algorithms, i.e., RFWD and OSPF, 50 times. Further, we statistically examine our results using Mann-Whitney U-test and linear regression analysis.

%\textbf{Internal validity}: To mitigate the threats posed by confounding factors, we compare the results obtained from executing DICES or the other algorithms, i.e., RFWD and OSPF, under the same parameter settings. As described in Section~\ref{subsec:param}, we present all the parameter settings of our experiments. Further, the DICES application and all our experiment data for any replication studies are available online~\cite{Material}. In addition, to mitigate potential biases in our experiment settings and data, we conducted our experiments based on inputs, i.e., network topology and data traffic profiles, from SES in the network domain.

%\textbf{Construct validity}: We mitigate the threats posed by unsuitable or ill-defined metrics, we use standard network quality metrics such as link utilization, packet loss, and delay. Further, we measured the execution time of DICES based on standard engineering practices.

\subsection{Threats to Validity}
%DICES is implemented using a specific open-source SDN framework (ONOS). Nevertheless and as described in Section~\ref{sec:approach}, DICES builds on general definitions of SDN concepts that do not depend on any specific SDN implementation. 
We evaluated DICES using both synthetic networks and an industrial IoT system. 
Since our current evaluation uses a network emulator (Mininet), future case studies and experiments on physical networks remain necessary for a more conclusive evaluation of DICES. In particular, there is the possibility that the physical network in our industrial case study system may sustain damage during natural disasters. DICES can operate properly as long as the underlying SDN provides accurate topology and traffic data. How accurate this data would be in the presence of network damage, and how one can counteract potential inaccuracies need to be further investigated. In addition, while motivated by IoT-enabled emergency management systems, DICES is a general congestion-control approach for SDN. Case studies in other domains, e.g., SDN-based data centers, are required in order to assess \hbox{the usefulness of DICES in a broader context.}

\section{Related Work}
\label{sec:related}

This section compares DICES with related work in the areas of communication protocols, SDN, IoT, self-adaptive systems and dynamic adaptive SBSE.

\textbf{Standard communication protocols} have been widely studied for resolving network congestion~\cite{Mathis:96,Alizadeh:10,Ferlin:16,He:16,Betzler:16}. For example, the TCP congestion control algorithm is prevalently used over the Internet and has been addressed by many prior research threads~\cite{Mathis:96,Alizadeh:10,Ferlin:16,He:16}. More recent work in this direction includes new application-layer protocols such as CoAP~\cite{CoAP:14} and its congestion control algorithm, CoCoA~\cite{Betzler:16}. These congestion control algorithms, in general, work by adjusting data transmission rates in an interconnected set of network hosts. In contrast, DICES  works by controlling the data flow paths and link weights in a network.

\textbf{SDN} has received considerable attention in the recent literature on networks. The problem of flow reconfiguration has been already studied for SDN with the objective of exploiting the additional flexibility offered by software~\cite{Brandt:16,Jin:14,Wang:16,Hong:13,Chiang:18,Agarwal:13,Gay:17,Huang:16}. %The SDN research strands that are most closely related to DICES are those studying online network traffic engineering which analyzes and regulates the behavior of data transmitted over a network. 
Chiang et al.~\cite{Chiang:18} formulate a new optimization problem to find optimal routing paths for group communication traffic. Gay et al.~\cite{Gay:17} propose a local search-based segment routing method for networks with unexpected failures. Huang et al.~\cite{Huang:16} present a dynamic routing algorithm to maximize network throughput under link-capacity and user-demand constraints. Agarwal et al.~\cite{Agarwal:13} present a single-objective linear programming method for improving network utilization. None of the above work strands account for the tradeoffs among the three objectives that DICES minimizes, i.e., maximum link utilization, number of link configurations, and delay. Further, unlike the above, DICES supports simultaneous dynamic control of data flow paths and link weights to both deal with the current congestion and also plan for handling future requests in a congestion-free manner. Finally, DICES is evaluated through a real case study on an emergency management system.

\textbf{IoT} may be realized through a variety of technologies and applied in many application domains~\cite{Al-Fuqaha:15,Taivalsaari:17}. The research topics related to IoT are numerous, e.g., data models to capture highly volatile IoT data~\cite{Moawad:15}, model-based code generation for heterogeneous things~\cite{Harrand:16}, model-based testing of IoT communications~\cite{Tappler:17}, IoT architectures~\cite{Koziolek:18,Potter:16}, and self-adaptive IoT systems~\cite{Iftikhar:17,Nascimento:17,Beal:17,Chen:16,Oteafy:17,Weyns:18}. Among these, an architecture-based adaptation framework by Weyns et al.~\cite{Weyns:18} is the most related to our work.  This prior work accounts for multiple quality of service criteria and represents them as quality constraints using state machines. In contrast, we formulate quality objectives as quantitative functions, thus enabling Pareto (tradeoff) analysis. Further, we use a multi-objective search algorithm to find practically acceptable solutions for dynamically reconfiguring an IoT system.
%To our knowledge, the problem of dynamically reconfiguring an IoT system in an unpredictable environment,  as tackled in our work, has not been studied before.

\textbf{Self-adaptive systems} have been studied in many domains~\cite{Krupitzer:18,Coker:15}. DICES relates to work on self-adaptation in the network domain, e.g., adaptive network anomaly detection~\cite{Ippoliti:16}, adaptive network monitoring~\cite{Bhuiyan:17,Anaya:14}, self-adaptive multiplex networking~\cite{Pournaras:18}, and network topology adaptation~\cite{Stein:16}. Among these, the most pertinent thread is by Stein et al.~\cite{Stein:16}, where the authors propose a topology adaptation model alongside a language to specify the adaptation logic of a set of network applications. This prior work aims to adapt a topology to a set of network applications, e.g., a video streaming source and a peer. In contrast, DICES adapts the network upon which the applications rely. Further, DICES uses multi-objective search to account for  optimization tradeoffs.
%DICES is further distinguished for the above work in that it performs the adapts a network at run-time.

\textbf{Dynamic adaptive SBSE}~\cite{Harman:12}, as noted in Section~\ref{sec:intro}, is the main research field upon whose principles we build.
%, applies the state-of-the-art of SBSE to find optimal system configurations at the compute step of the feedback-loop control in Figure~\ref{fig:overview} for dynamic self-adaptation~\cite{Harman:12}. 
Prior research in this field has employed search for various purposes, e.g., improving the design and architecture of self-adaptive systems~\cite{Menasce:11,Andrade:13,Ramirez:10} and configuring such systems~\cite{Ramirez:09,Zoghi:16}. To our knowledge, we are the first to have addressed the problem of congestion control in the context of dynamic adaptive SBSE.
%in networking applications. 
%Among the existing threads of research, the most related to DICES is the work by Zoghi et al.~\cite{Zoghi:16} where the authors present a method to configure adaptive systems in networked environments. The method consists of several steps including the elicitation of stakeholders' adaptation goals, the transformation of these goals into control points, and the activation of the transformed control points using search-based algorithms.  DICES accounts for multiple objectives using multi-objective search and resolves congestion in a dynamic adaptive manner. In addition, we evaluate DICES using synthetic networks for controlled experiments as well as the EMS network for conducting experiments under real-world conditions. We further demonstrate that recent advances in SDN and dynamic adaptive SBSE make our work promising research.
% !TEX root =  paper.tex

\section{Conclusions}
\label{sec:conclusions}

We developed a search-based approach, named DICES, to dynamically mitigate network congestion in IoT systems via network reconfiguration. Our approach is realized through a control feedback loop, whereby the traffic on an IoT network is periodically monitored and corrective action is taken at run-time when congestion is detected. The corrective action to take (i.e., the reconfiguration) is computed using a multi-objective search algorithm that simultaneously minimizes: (1)~the maximum link utilization across all the links in the network, (2)~the number of link updates for reconfiguration, and (3)~the overall data transmission delays. %In a nutshell, the approach works by discouraging the use of highly utilized and low-speed links, and encouraging the use of lowly utilized and high-speed links.
We evaluated DICES on a number of synthetic networks as well as an industrial IoT-enabled emergency management system. The results indicate that DICES is able to efficiently and effectively adapt an IoT network to resolve congestion. Further, compared to two common data forwarding algorithms which we use as baselines, DICES yields data transmission rates that are at least 3 times faster while reducing data loss by at least 70\%.

%\begin{sloppypar}
%For future work, we plan to extend DICES by accounting for: (1)~link and switch failures and (2)~the policies (e.g., cost-containment policies) that govern the use of terrestrial and satellite telecommunication networks. 
%When dealing with fast-changing network loads (i.e., when $Q_{i+1}$ is drastically different from $Q_i$), the flows computed by DICES to optimize network usage based on $Q_i$  may not be optimized for $Q_{i+1}$. We intend to address this issue in future by using prediction models to anticipate $Q_{i+1}$ earlier and use it in the computation of the optimized flows for the next step.
%In the longer term, we would like to further validate DICES by deploying it as an integrated component of the emergency management system in our industrial case study (in-situ deployment).
%\end{sloppypar}
%Beyond emergency management systems, we will explore an alternative simulation method, which does not rely on wall-clock time, to facilitate the development of dynamic adaptive configuration techniques for IoT network systems.

\begin{acks}
This project has received funding from SES, the Luxembourg National Research Fund under the grant C-16PPP/IS/11270448 and the European Research Council (ERC) under the European Union's Horizon 2020 research and innovation programme (grant agreement No 694277).
\end{acks}

\bibliographystyle{ACM-Reference-Format}
\balance
\bibliography{paper}

\end{document}